\documentclass[12pt]{article}
\pdfoutput=1
\usepackage{graphicx}
\usepackage{amsmath,amsfonts,amssymb,stmaryrd,latexsym}
\usepackage{url,subcaption,comment}
\usepackage{hyperref}
\usepackage{xcolor}
\hypersetup{
    colorlinks=blue,       
    linkcolor=purple,   
    citecolor=blue,        
    filecolor=magenta,      
    urlcolor=black           
}
\usepackage{cite}
\usepackage{dcolumn}
\usepackage{bm}
\usepackage{booktabs ,ragged2e}
\newcommand{\be}{\begin{equation}}
\newcommand{\ee}{\end{equation}}
\newcommand{\bea}{\begin{eqnarray}}
\newcommand{\eea}{\end{eqnarray}}
\newcommand{\bear}{\begin{eqnarray}}
\newcommand{\eear}{\end{eqnarray}}
\newcommand{\ba}{\begin{array}}
\newcommand{\ea}{\end{array}}

\newcommand{\taup}{{\tau'}}
\newcommand{\atau}{{a_\tau}}
\newcommand{\phase}{{\beta}}

\begin{document}

\baselineskip=18pt \pagestyle{plain} \setcounter{page}{1}

\vspace*{-1.cm}

\noindent \makebox[11.9cm][l]{\small \hspace*{-.2cm} }{\small Fermilab-Pub-23-151-T}  \\  [-1mm]

\begin{center}

{\Large \bf    
Vectorlike leptons and long-lived  bosons at the LHC
} \\ [1cm]

{\normalsize \bf Elias Bernreuther and Bogdan A. Dobrescu \\ [4mm]
{\small {\it Particle Theory Department, Fermilab, Batavia, IL 60510, USA     }}\\ [14mm]
}

\center{August 29, 2023}

\end{center}

\vspace*{0.8cm}

\begin{abstract} \normalsize
We study a renormalizable theory that includes a heavy weak-singlet vectorlike lepton, $\tau^\prime$, that decays into a pseudoscalar boson and a tau lepton. We show that this can be the dominant decay mode of $\tau^\prime$ provided the pseudoscalar couplings deviate from the case of a Nambu-Goldstone boson. For a range of parameters, the pseudoscalar is long-lived, and mainly decays into a photon pair at one loop, or into a tau pair at tree level. Electroweak production of $\tau^{\prime +} \tau^{\prime -} $ leads to a rich phenomenology at colliders, including signals with many taus or photons. We analyze in detail the LHC signal involving two prompt taus and two displaced photon pairs.
Particularly interesting is the case where the pseudoscalar has a decay length of a few meters and thus 
would typically deposit energy in the muon chambers of  the CMS or ATLAS detectors. 
\end{abstract}

\newpage

\vspace*{1cm}

\renewcommand{\contentsname}{\normalsize\large Contents}
{ \hypersetup{linktocpage} 
\tableofcontents
\hypersetup{linkcolor=red} 
}

\newpage
\section{Introduction} \setcounter{equation}{0}
\label{sec:intro}

All the Standard Model (SM) fermions are chiral, as their left-handed components are weak doublets, while the right-handed components are singlets.
LHC experiments have essentially ruled out the existence of additional chiral fermions. Thus, if new elementary fermions exist, then they must be vectorlike ({\it i.e.}, non-chiral)  with respect to the SM gauge interactions. Searches for vectorlike quarks (Vquarks) at the LHC have been performed for almost a decade, and have set a lower limit on their masses of about 1.5 TeV \cite{ATLAS:2018ziw}. By contrast, searches for vectorlike leptons (Vleptons) are only now ramping up, due to the much smaller production cross section for color-singlet particles.

Pair production of Vleptons has been searched for under the hypothesis that the Vlepton decays via mass-mixing with the $\tau$, into $W\nu$, $Z \tau$, $h^0 \tau$  \cite{CMS:2022nty} (see also \cite{CMS:2019hsm},\cite{Muse:2022xgu}). We will refer to these as the standard decay modes of a Vlepton. Theoretical studies of these standard modes at the LHC can be found in Ref.~\cite{Kumar:2015tna,Falkowski:2013jya}.
An alternative decay hypothesis, searched for by CMS  \cite{CMS:2022pvz},  is the Vlepton decay  into $\tau jj$ through an off-shell leptoquark (similar exotic decays of Vquarks have been studied in \cite{Dobrescu:2016pda}).

Here we study the possible Vlepton decay into a new spin-0 particle and a $\tau$. We focus on a model in which 
a gauge-singlet complex scalar has Yukawa interactions with an $SU(2)_W$-singlet fermion whose left- and right-handed components have the same gauge charges as the SM right-handed $\tau$.
The mass mixing of the Vlepton with the SM third-generation charged lepton leads to a heavy physical lepton labelled $\tau^\prime$ and 
a lighter physical lepton identified with the well-known $\tau$ lepton. 
We consider the case where the complex scalar has a VEV, and its CP-odd component ($a_\tau$) is lighter than 
$\tau^\prime$. The branching fraction for $\tau^\prime \to \tau  a_\tau  $ competes with the standard decay modes mentioned above. 
If the Vlepton mass is due to the scalar VEV, then $a_\tau$ is a pseudo-Nambu-Goldstone and $\tau^\prime \to \tau  a_\tau  $ has a branching fraction that depends predominantly on the VEV, and is typically smaller than the branching fractions  for the standard modes. If, however, the Vlepton mass arises from Lagrangian terms that do not depend on the VEV, then the $\tau^\prime \to \tau  a_\tau  $ branching fraction may be by far the largest.

Pair production of $\tau^\prime$ followed by $\tau^\prime \to \tau a_\tau$ decays leads to various interesting signals at the LHC. Here we analyze the signature arising when each of the two $a_\tau$  particles decays into two photons. These $a_\tau \to \gamma \gamma$ decays occur at one loop, with the $\tau^\prime$ running in the loop. For a $\tau^\prime$ mass of order 1 TeV, an $a_\tau$ mass at the GeV scale, and Yukawa couplings of about 0.1, the typical decay length of $a_\tau$ in the lab frame is in the range of a few meters to roughly 30 m.
As a result, the signal at CMS or ATLAS would be two prompt $\tau$'s and two energy deposits in the muon chambers. The latter benefit from a highly suppressed background, allowing for experimental sensitivity to $m_{\tau^\prime}$ of up to about 1 TeV.  Existing CMS and ATLAS searches involving long-lived particles (LLPs) which lead to deposits in the muon systems \cite{CMS:2021juv, ATLAS:2018tup, ATLAS:2019jcm} have shown better sensitivity to longer lifetimes than the well-established searches involving the calorimeters or the tracker \cite{Alimena:2019zri}.

The model studied here is renormalizable and includes only two simple fields beyond the SM, so it is interesting in its own right. Nevertheless, this model may be the low energy manifestation of a richer underlying theory. An example is a theory of quark and lepton compositeness \cite{Dobrescu:2021fny} in which vectorlike fermions appear as confined states of a strongly coupled chiral interaction, 
and complex spin-0 fields with VEVs appear as bound states of two composite fermions. The lightest 
composite vectorlike fermion in that theory is a weak-singlet Vlepton with the same quantum numbers as the one discussed here.
Another example of theories that automatically include the particles considered here (while additional new particles could be too heavy for production at the LHC) are $U(1)$ extensions of the SM gauge group in which the weak-singlet Vlepton is an anomalon, {\it i.e.}, a fermion required to cancel the gauge anomalies\footnote{The fermions we study here are vectorlike under the SM gauge group, but may be chiral under some new gauge groups, as is the case for anomalons.}, and the spin-0 particle is part of the symmetry breaking sector \cite{Dobrescu:2014fca}.
Electroweak baryogenesis in a model 
that predicts LHC signals with a Vlepton decaying into a scalar and a tau is studied in \cite{Bell:2019mbn}.

This paper is organized as follows.
In Section 2, after we present the model, we derive the couplings and decay widths of the new particles. Production and various signals at the LHC are discussed in Section 3. The case of a long-lived $a_\tau$ produced in Vlepton decays at the LHC is analyzed in detail in Section 4. Our conclusions are summarized in Section 5.

\bigskip

\section{Weak-singlet Vlepton plus a complex scalar}  
\setcounter{equation}{0}
\label{sec:model}

Consider an extension of the SM that includes a Dirac fermion $\cal E$ and a complex scalar $\phi$, with a Yukawa interaction $\phi \, \overline {\cal E }_L  {\cal E}_R $.
The complex scalar has a potential with the minimum at $\langle \phi \rangle = v_\phi > 0$, and can be written as
\be
\phi =  \left( v_\phi + \frac{1}{\sqrt{2} } \, \varphi_\tau  \right)  e^{i a_\tau/ ( \sqrt{2} \, v_\phi )}     ~~.
\label{eq:phi}
\ee
Here $a_\tau$ is a CP-odd scalar, and $\varphi_\tau$ is a CP-even scalar. We assume that the scalar potential includes some terms 
with small coefficients that break explicitly the global $U(1)$ symmetry associated with $\phi$. It is thus natural that the mass of $a_\tau$ 
satisfies $M_a \ll v_\phi$, while the mass of $\varphi_\tau$ is roughly of the order of $v_\phi$. 

We focus on the case where $\cal E$ is a singlet under the $SU(3)_c \times SU(2)_W$ gauge group, and both its left-handed and right-handed components carry hypercharge $-1$. Thus, $\cal E$ is a weak-singlet vectorlike lepton, which can be produced in pairs
at the LHC through its couplings to the $Z$ and the photon. 
Since $\cal E_R$ has the same charges as the SM right-handed weak-singlet leptons, $ e_R^j$, $j = 1,2,3$, gauge invariance allows Yukawa couplings to the SM Higgs doublet:  $H  \,  \overline  \ell^j_L \,  {\cal E }_R $, where $\ell^j_L = (\nu^j_L , e^j_L) $ is the SM lepton doublet of the $j$th generation, with $j = 1,2,3$. 
Given the strong experimental limits on violation of lepton universality from pion decays, the mixing of the Vlepton with the SM electron and muon fields must be very small, of order $10^{-3}$ or below \cite{Endo:2014hza}. The constraints from lepton universality on the flavor-diagonal tau interactions are less stringent. 
Nevertheless, lepton universality violation of the tau in conjungation with off-diagonal Higgs Yukawa couplings involving a tau field and an electron or muon field are constrained by limits on lepton-flavor violating processes such as  $\tau \to e \gamma$ and $\tau \to \mu \gamma$. 
This indicates the existence of a global symmetry that prevents the mixing of the electron and muon with the tau, while allowing mixing of the tau with the vectorlike lepton.
An example is an $U(2)_L\times U(2)_R$ global symmetry acting on the electron and muon fields, or alternatively on the tau and $\cal E$ fields.

As only one linear combination of $\cal E_R$ and $e_R^3$ couples to $\ell_L^3$, we can use a global $U(2)_R$ transformation to redefine the right-handed fields such that the most general Higgs Yukawa interaction of  the third generation lepton fields and $\cal E$ is
\be
-  y_3  \, H  \,  \overline  \ell^3_L   \,  e^3_R   +  {\rm H. c.}       ~~,
\label{eq:yukH}
\ee
where $y_3 > 0 $ is a Yukawa coupling.  In this basis for $\cal E_R$ and $e_R^3$, the  Yukawa interactions of $\phi$ take the form
\be
-  \phi \, \overline {\cal E }_L  \left(  y_{\cal E } \, e^{i\phase_{\cal E }}  \,   {\cal E}_R 
+ y_o \, e^{i \phase_o}  \, e_R^3 \right)  +  {\rm H. c.}     ~~,
\label{eq:yukE}
\ee
where $y_{{\cal E } }, y_o > 0 $ and the complex phases satisfy $ 0 \leq \phase_{\cal E }, \phase_o <  2 \pi$.
Mass terms that link ${\cal E }_L $ to the weak-singlet fermions may also be present, and are generically of the form
\be
- m_{{\cal E } {\cal E }}\, \overline {\cal E }_L \,   {\cal E }_R
- m_{{\cal E } 3}\, \overline {\cal E }_L \,  e_R^3 +  {\rm H. c.}    
\label{eq:mE}
\ee
 These Lagrangian terms and the Yukawa interactions (\ref{eq:yukE}) give the effective mass terms  
 $-  \overline {\cal E }_L \,   ( m_{\cal E } \,  {\cal E }_R +  m_o \, e^3_R ) $,  where 
\bear
m_{{\cal E } } = m_{\cal E E} + y_{\cal E }  \, e^{i\phase_{\cal E }}  \, v_\phi  > 0 ~~,
\nonumber
\\  [-3mm]
\\ [-3mm]
m_o = m_{{\cal E} 3} + y_o \, e^{i\phase_o}  \, v_\phi  > 0 ~~.
\nonumber
\eear
The above choice of the $m_{{\cal E } } $ and $m_o$ phases is an outcome of a field redefinition of ${\cal E }_R$, $e^3_R$ and $\ell_L^3$
that does not affect (\ref{eq:yukH}) or (\ref{eq:yukE}).

Putting together all the Lagrangian terms discussed so far, and expanding the complex scalar \eqref{eq:phi} in $a_\tau$,
the most general Yukawa couplings and mass terms involving the Vlepton and the SM tau fields can be written as
\be
- \left( \overline e^3_L \, , \;  \overline  {\cal E }_L  \right)   
 \left( \ba{cc}   y_3  \left( v_H +   { \displaystyle \frac{ h^0}{\sqrt{2}} }   \right)    &    0 
    \\  [4mm]  
 m_o  +  { \displaystyle \frac{ y_o  \, e^{i \phase_o} }{\sqrt{2} } }   \left( \varphi_\tau  + i a_\tau \right)  \;
 & \;  m_{\cal E } +  { \displaystyle \frac{ y_{\cal E }  \, e^{i \phase_{\cal E }} }{\sqrt{2} } } 
  \left( \varphi_\tau  + i a_\tau \right)   \ea  \right)  
 \left( \ba{c}   e^3_R  \\ [3mm]   {\cal E }_R  \ea  \right)   + {\rm H.c.} ~.
 \label{eq:matrix}
\ee
Here $h^0$ is the SM Higgs boson and $v_H \approx 174 $  GeV is  the weak scale.
Note that all parameters that appear in these Lagrangian terms  
are positive.  The $2\times 2$ mass matrix can be diagonalized by the $SU(2)_L\times SU(2)_R$ transformation 
\be
\left( \ba{c}   e^3_{L,R}  \\  {\cal E }_{L,R}   \ea  \right)   = 
\left( \ba{cc}  c_{L,R}   &   s_{L,R}   \\ [2mm]    -s_{L,R}   &   c_{L,R}   \ea  \right)  
\left( \ba{c}   \tau_{L,R}  \\  \tau^\prime_{L,R}    \ea  \right)    ~~.
\ee
where $\tau^\prime_{L,R} $ are the  left- and right-handed components of a heavy charged lepton, and  
$\tau_{L,R} $ are the left- and right-handed components of the observed tau. The notation used here is 
$c_{L,R} \equiv \cos \theta_{L,R}$,  $s_{L,R} \equiv \sin \theta_{L,R}$, where the mixing angles 
$\theta_L$ and $\theta_R$ are real parameters in the $[-\pi,\pi)$ range.

\subsection{Couplings of the Vlepton}

The mixing angles $\theta_L$ and 
$\theta_R$ are functions of the three input parameters from the mass matrix:
$m_{\cal E } $, $m_o$, and $y_3  v_H$. There are four relations between these quantities, 
because the mass matrix in the physical basis must satisfy
\be \hspace{-0.51cm}
\left( \ba{cc}  \!\! c_L c_R \,  y_3 v_H  + s_L ( s_R \, m_{\cal E } \! - \! c_R  \, m_o ) 
& c_L s_R \,  y_3 v_H  - c_L ( s_R \, m_{\cal E } \! + \! s_R  \, m_o )       \!\! 
 \\  [4mm]  
\! s_L c_R \,  y_3 v_H  - c_L ( s_R \, m_{\cal E } \! - \! c_R  \, m_o ) 
&  s_L s_R \,  y_3 v_H  + c_L ( c_R \, m_{\cal E } \! + \! s_R  \, m_o ) \!\! \ea  \right)  
= \left( \!\! \ba{cc}   m_\tau  &    0     \\  [4mm]  0  &  m_{\tau'}    \ea  \!\!  \right)  ~,
\ee 
where $m_\tau$ is the measured mass of the tau particle, and $m_{\tau'}$ is the mass of the 
heavy lepton that will be searched for in collider experiments.
Solving for the three input parameters, we find
\bear
&& m_{\cal E }  = m_{\tau'}   \frac{c_R}{c_L}    ~~,
\nonumber \\ [2mm]
&& m_o = m_{\tau'} \frac{s_R}{c_L} - m_{\tau} \frac{s_L}{c_R}   ~~,
\label{eq:mome}
\\ [2mm]
&& y_3 v_H   = m_{\tau}   \frac{c_L}{c_R}   ~~,
\nonumber
\eear
while the fourth equation leads to a relation between $\theta_R$ and $\theta_L$ \cite{Dobrescu:2009vz}, which is useful to write as: 
\be
 \tan\theta_R  =   \frac{m_{\tau'} }{m_{\tau} }    \,    \tan\theta_L     ~~.
\label{eq:tRtL}
\ee

The off-diagonal couplings of the spin-0 particles $a_\tau$ and $\varphi_\tau$ to the physical fermions $\tau$ and $\tau^\prime$ are
\be
\frac{ \varphi_\tau + i a_\tau}{\sqrt{2} } \;   \overline{\tau}   \left(  c_L \,  {\cal Y}_o   \, e^{-i \phase_o }  P_L + s_L \,  {\cal Y}_{\cal E }   \, e^{i \phase_{\cal E } }  P_R \right)   \tau^\prime + {\rm H.c.}  ~~,
\ee
where $P_{L,R} = (1 \mp \gamma_5)/2$, and the coupling constants introduced here are defined by
\bear
&&   {\cal Y}_o  = y_o  \, c_R -   y_{\cal E } \, s_R \, e^{i (\phase_o - \phase_{\cal E } )}    ~~,
\nonumber \\  [-1mm]
\label{eq:gLgR}
 \\  [-1mm]
&&   {\cal Y}_{\cal E }   =   y_{\cal E }  \, c_R    +  y_o \,  s_R  \, e^{i (\phase_o - \phase_{\cal E } )}    ~~.
\nonumber
\eear
The diagonal couplings of $a_\tau $ and $\varphi_\tau$  to tau's or tau-primes  are given by 
\bear
&& - c_L \, \frac{ \varphi_\tau + i a_\tau}{\sqrt{2} } \;  \overline{\taup} \left(  y_\mathcal{E} \, c_R  \, e^{i \phase_\mathcal{E}}  +  y_o \, s_R  \, e^{i \phase_o} \right) P_R \, \taup + \mathrm{H.c.} \; ,
\nonumber \\  [-1mm]
\label{eq:diagCouplings}
 \\  [-1mm]
&&
s_L \, \frac{ \varphi_\tau + i a_\tau}{\sqrt{2} } \;   \overline{\tau} \left( y_o  \, c_R  \, e^{i \phase_o} -   y_\mathcal{E}  \, s_R  \, e^{i \phase_\mathcal{E}} \right) P_R \, \tau + \mathrm{H.c.} \; 
\nonumber
\eear
There is also an off-diagonal coupling of the SM Higgs boson to $\tau$ and $\tau^\prime$: 
\be
- \frac{y_3  }{\sqrt{2} } \;  h^0 \,  \overline{\tau}   \left( s_L c_R    P_L + c_L s_R  P_R   \rule{0mm}{3.9mm} \right) \tau^\prime
 + {\rm H.c.} 
\ee

The  $W$ boson interacts with a $\tau^\prime$ or a $\tau$ and the tau neutrino as follows:
\be
\frac{g}{\sqrt{2}}   W^-_\mu \,  \overline \nu_{\tau}  \gamma^\mu  P_L \left( c_L \,  \tau  + s_L \, \tau^\prime \right)
+ {\rm H.c.} 
\ee
Since the only partial decay width of $W$ affected by the vectorlike lepton is that 
 into $\tau \nu_\tau$, the ratio of branching fractions $R_W(\tau/\mu) = B(W \to \tau \nu_\mu)/B(W \to \mu \nu_\mu)$ is predicted to be  $R_W(\tau/\mu) =  c_L^2$. The recent measurement of this ratio \cite{ATLAS:2020xea}, 
 $R_W(\tau/\mu) = 0.992 \pm 0.013$, implies $s_L^2 = 0.008 \pm 0.013$, so that it is justified to expand 
 in $s_L^2$.
There is also an off-diagonal interaction of the $Z$ boson to a $\tau^\prime$ and a $\tau$,
\be
- \frac{c_L s_L  \, g}{\cos\theta_W} \,   Z_\mu \, 
\overline \tau_L   \gamma^\mu  \tau_L^\prime  + {\rm H.c.}    ~~,
\ee
which together with the other off-diagonal interactions discussed above 
induces 2-body decay modes for the heavy lepton.

The diagonal interactions of the $Z$ boson to the $\tau$ and $\tau^\prime$  leptons are 
\be
\frac{g}{ 2 \cos\theta_W}   Z_\mu \,   \left[ \overline \tau   \gamma^\mu  
\left( - \frac{c_L^2}{2} P_L + \sin^2\theta_W \right) \tau
+ \overline \tau^\prime    \gamma^\mu  
\left( - \frac{s_L^2}{2} P_L + \sin^2\theta_W \right) \tau^\prime  \right]    ~~.
\ee
This implies that the partial width for $Z \to \tau^+ \tau^-$ is reduced compared to the SM such that the ratio $R_Z(\tau/\ell)$
of $Z$ branching fractions into taus and into $\ell^+\ell^-$ (which is the average of the $e^+e^-$ and $\mu^+\mu^-$ branching fractions) deviates from the lepton-universality prediction of 1:
\be
R_Z(\tau/\ell) \equiv\frac{B(Z \to  \tau^+ \tau^-)}{B(Z \to \ell^+\ell^-)} = 
1 - 2 s_L^2  \, \frac{ 1 - 2  \sin^2\theta_W }{ 1 - 4 \sin^2\theta_W + 8 \sin^4\theta_W } ~~.
\label{eq:RZ}
\ee
Using $B(Z \to e^+ e^-) = (3.3632 \pm 0.0042)\%$ and $B(Z \to \mu^+ \mu^-) = (3.3662 \pm 0.0066)\%$ \cite{ParticleDataGroup:2022pth},
we obtain $B(Z \to \ell^+\ell^-) = (3.3647 \pm 0.0039)\%$. The branching fraction into taus has a slightly larger 
central value, $B(Z \to \tau^+ \tau^-) = (3.3696 \pm 0.0083)\%$, so that $R_Z(\tau/\ell) = 1.0015 \pm 0.0027$.
Comparing this value with the theoretical prediction (\ref{eq:RZ}), we find 
$s_L^2 < 1.4 \times 10^{-3}$ at the 95\% CL. This upper limit  is more stringent by an order of magnitude than the one set by the $W$ measurement mentioned earlier, and it  implies  the following 95\% CL upper limit on $s_L$: 
\be
s_L < 0.037   ~~.
\ee

\subsection{Decays of the Vlepton}

Due to the off-diagonal couplings derived above, the physical heavy lepton $\tau^\prime$ has decay modes into a tau 
and a neutral massive boson ($a_\tau$, $\varphi_\tau$, $h^0$, $Z$), as well as into $W\nu$.
The tree-level width for the vectorlike lepton decay into $\tau a_\tau$ is given by 
\be
\Gamma \left( \tau^\prime \to \tau  a_\tau  \right) =\frac{ m_{\tau'}}{64 \pi}   \left(   c_L^2 \,  | {\cal Y}_o |^2 + s_L^2 \, |{\cal Y}_{\cal E } |^2 \right) \left( 1 -  \frac{M_{a}^2}{m_{\tau'}^2}    \right)^{\! 2}    ~~.
\ee
The width for $ \tau^\prime \to \tau  \varphi_\tau $ is obtained from the above result 
by replacing  $M_{a}$ with the $\varphi_\tau$ mass $M_\varphi$, due to the different phase-space.

The width into a tau and a Higgs boson is 
\be
\Gamma \left( \tau^\prime \to \tau  h^0  \right) = \frac{ m_{\tau'}}{64 \pi}  \; y_3^2  \, \left( c_L^2 s_R^2  +  s_L^2 c_R^2 \right) 
\left( 1 -  \frac{M_h^2}{m_{\tau'}^2}   \right)^{\! 2}    ~~.
\ee
Eq.~(\ref{eq:tRtL}) and the last equation in (\ref{eq:mome}) imply 
\be
y_3^2  \, \left( c_L^2 s_R^2  +  s_L^2 c_R^2 \right) = \frac{ c_L^2 s_L^2 }{ v_H^2}  \left(  m_{\tau'}^2 +  m_{\tau}^2 \right)  ~~,
\ee
so that to leading order in $s_L^2$ and in $m_\tau^2/m_{\tau'}^2$ the $\tau^\prime \to \tau  h^0$ width can be written as
\be
\Gamma \left( \tau^\prime \to \tau  h^0  \right) = \frac{ s_L^2 m_{\tau'}^3}{64 \pi v_H^2}  \left( 1 -  \frac{M_h^2}{m_{\tau'}^2}   \right)^{\! 2}    ~~.
\ee

The remaining 2-body decay widths of $\tau'$ into SM particles are, again to leading order in $s_L^2$ and in $m_\tau^2/m_{\tau'}^2$, 
given by
\bear
&& \Gamma \left( \tau^\prime \to \tau  Z^0 \right) =  \frac{ s_L^2 m_{\tau'}^3}{64 \pi v_H^2}  \, \left( 1 -  3 \frac{M_Z^4}{m_{\tau'}^4} + 2 \frac{M_Z^6}{m_{\tau'}^6}   \right)   ~~,
\nonumber \\  [-1mm]
\label{eq:WZwidths}
 \\  [-1mm]
&& \Gamma \left( \tau^\prime \to \nu_\tau  W^- \right) 
= \frac{ s_L^2 m_{\tau'}^3}{32 \pi v_H^2}  \, \left( 1 - 3 \frac{M_W^4}{m_{\tau'}^4} + 2  \frac{M^6_W}{m_{\tau'}^6} \right)  ~~.
\nonumber
\eear
For $m_{\tau^\prime} \gg M_h$, the widths for $ \tau^\prime \to \tau  h^0,  \, \tau  Z^0, \, \nu  W$ are in the ratio $1:1:2$, as 
expected \cite{Han:2003wu} based on the Goldstone equivalence theorem \cite{Chanowitz:1985hj}. 

The ratio between the decay width for $\tau^\prime \to \tau  a_\tau$ and the sum of the SM widths ($\tau^\prime \to \tau h^0,  \tau Z^0, \nu_\tau W^-$) is given in the limit  $s_L^2 \ll1$ by
\be
R_\tau  \equiv  \frac{\Gamma \left( \tau^\prime \to \tau  a_\tau  \right) }{\Gamma \left( \tau^\prime \to {\rm SM}  \right) }
= \frac{  v_H^2 \left( |{\cal Y}_o |^2  / s_L^2 + |{\cal Y}_{\cal E }  |^2 \right) }{4m_{\tau'}^2  - 2 M_h^2  + O(M_h^4/m_{\tau'}^2)}  
 \left( 1 -  \frac{M_{a}^2}{m_{\tau'}^2}    \right)^{\! 2}    ~~.
\label{eq:Rtau}
\ee
Within various regions of parameter space $R_\tau \gg 1$ so that $\tau^\prime \to \tau  a_\tau $ is the dominant decay mode.
For illustration, if $y_o \approx y_{\cal E } \approx 1$,
$\phase_o - \phase_{\cal E }  \approx \pi$, and $m_{\tau'}^2 \gg M_{h}^2, M_a^2$,  
then $R_\tau \approx  (c_R + c_R^2/s_R)^2  \, v_H^2/ (4 m_\tau^2) \, $; thus
$\tan\theta_R$ in the intervals  $(-3.2, -1.7) \cup (-0.8,5.7)$ yields  $R_\tau \gtrsim 100$.

\subsection{Nambu-Goldstone limit}

In the $m_{\cal E E} = m_{{\cal E}3} =0$ limit the Yukawa interactions have a global $U(1)$ symmetry that is spontaneously broken by the $\phi$ VEV. In that limit, $a_\tau$ is a pseudo-Nambu-Goldstone boson, with its small mass ($M_a \ll m_{\tau^\prime}$) originating from $U(1)$ breaking terms in the scalar potential.
The $\phase_{\cal E }$ and $\phase_o$ phases in that case can be set to 0 by a redefinition of the 
 ${\cal E }_R$  and  ${\cal E }_L$ fields. Thus, $m_{{\cal E } } =  y_{\cal E }  \, v_\phi$ and $m_o = y_o \, v_\phi$, 
so from (\ref{eq:mome}) and (\ref{eq:tRtL}) follows that Yukawa couplings are 
\bear
&& 
y_{\cal E} =  \frac{c_R \,  m_{\tau^\prime} }{ c_L \, v_\phi }  ~~,
\nonumber \\  [-1mm]
\label{eq:gLgR}
 \\  [-1mm]
&&  y_o = s_L c_R \, \frac{m_{\tau^\prime}^2 - m_\tau^2}{v_\phi \, m_\tau}  ~~.
\nonumber
\eear
As a result, the ratio of the $\tau^\prime \to \tau a_\tau$ width and the sum of $\tau'$  widths into SM particles is given, in the Nambu-Goldstone case, by
\be
R_\tau =   \frac{ v_H^2 }{ 4 v_\phi^2 } \;  \left(  1 + \frac{M_h^2}{2 m_{\tau^\prime}^2}  + \, O(M_h^4/m_{\tau'}^4) \right)  ~~
\ee
to leading order in the small quantities $s_L^2$ and $(m_\tau /m_{\tau^\prime})^2$.
This shows that a measurement of $R_\tau$ would represent a measurement of the $\phi$ VEV in the 
case where the $a_\tau$ pseudoscalar has approximately the couplings of a Nambu-Goldstone boson. 

From the first two equations in (\ref{eq:mome}) follows that 
\be
v_\phi^2 = \frac{ m_{\tau^\prime}^2 }{ y_{\cal E }^2 + y_o^2}   
\left[1 + O(s_L^2) + O\left(m_\tau^2 /m_{\tau^\prime}^2 \right) \right]  ~~,
\ee
so we can write 
\be
R_\tau 
\approx 0.19 \,   \left( y_{\cal E }^2 + y_o^2 \right) \left( \frac{200 \; {\rm GeV} }{m_{\tau^\prime}} \right)^{\! 2}
 \label{eq:RtauGB}
\ee
for $m_{\tau^\prime}^2  \gg M_h^2$.
Given that the Yukawa couplings $y_{\cal E }$ and $y_o$ cannot be larger than about 2 without the theory becoming nonperturbative at a scale near $v_\phi$, (\ref{eq:RtauGB}) implies an upper limit  on $R_\tau$, which  decreases from 1.5 at  $m_{\tau^\prime} = 200$ GeV to 0.061 at  $m_{\tau^\prime} = 1$ TeV.
Note that the exotic branching fraction satisfies $B(\tau^\prime \to \tau a_\tau) = R_\tau /(1 + R_\tau)$, so it decreases from 60\% to 6\% when $m_{\tau^\prime}$ varies in the 0.2 -- 1 TeV range (for $y_\mathcal{E}=y_o=2$). For the rest of the paper we assume that $B(\tau^\prime \to \tau a_\tau) \gtrsim 70~\%$, which happens for certain ranges of parameters away from the Nambu-Goldstone-boson limit, as discussed after Eq.~\eqref{eq:Rtau}.

\subsection{Decay modes of the spin-0 particles}  
\label{sec:length}

The diagonal couplings (\ref{eq:diagCouplings}) of the spin-0 particles to $\tau^\prime$ can be rewritten as
\be
- \frac{1}{\sqrt{2} } \;  \left( {\rm Re} \, y_{\taup} \; a_\tau + {\rm Im} \, y_{\taup} \; \varphi_\tau  \right)  \, i \,  \overline{\taup} \gamma_5 \taup 
+  \frac{1}{\sqrt{2} } \;  \left( {\rm Im} \, y_{\taup} \;a_\tau - {\rm Re} \, y_{\taup} \; \varphi_\tau  \right)  \; \overline{\taup}  \taup 
~~ , 
\label{eq:pseudo}
\ee
where the effective Yukawa coupling is 
\be
y_{\taup} =  y_o \, s_R  \, e^{i \phase_o}  + y_\mathcal{E} \, c_R  \, e^{i \phase_\mathcal{E}}  ~~.
\ee
Note that in the limit where $y_{\taup}$ is a real number $a_\tau$ has only the pseudoscalar coupling to $\tau^\prime$ and is a CP-odd particle, while $\varphi_\tau$ has only the scalar coupling and is CP-even.
Given that the phases $\phase_\mathcal{E} , \phase_o$ are in the $[0,2\pi)$ interval, the CP symmetry is generically violated. Thus, $a_\tau$ and $\varphi_\tau$ have both the pseudoscalar and scalar couplings to the fermions even when the tree-level potential for $\phi$ is CP invariant. However, the CP-violating terms in (\ref{eq:pseudo}) are typically suppressed, so for simplicity we will refer to $a_\tau$ as a ``pseudoscalar" and to $\varphi_\tau$ as a ``scalar".

The diagonal couplings (\ref{eq:diagCouplings}) of the spin-0 particles to $\tau$ leptons are suppressed by the small mixing $s_L$, and 
can be written in the same form as (\ref{eq:pseudo}) with $y_{\taup}$ replaced by the effective Yukawa coupling
\be
y_{\tau}  =  s_L \left( y_o  \, c_R  \, e^{i \phase_o} -   y_\mathcal{E}  \, s_R  \, e^{i \phase_\mathcal{E}} \right) ~~.
\ee
For $2 m_\tau < M_a < m_\taup + m_\tau$, the process $a_\tau \to \tau^+ \tau^-$ is the only 2-body decay at tree level of $a_\tau$.
Its width is
\be
\Gamma \left( a_\tau \to \tau^+ \tau^- \right)  =  \frac{ \left| y_{\tau} \right|^2 }{16 \pi } \, M_a \left( 1 - 4 \frac{m_\tau^2}{M_a^2} \right)^{3/2}   ~~.
\ee

The interactions in (\ref{eq:pseudo}) and the analogous ones involving taus induce at one loop the decay of $a_\tau$ into a pair of photons (and also into $Z\gamma$ for $M_a > M_Z$).
Both a $\taup$ loop and a $\tau$ loop contribute to this process, with the dominant contribution depending mainly on the relative size of $s_L$
and $m_\tau/m_\taup$, but also on $ \phase_o$ and $\phase_\mathcal{E} $.
If the $\tau$ loop dominates and $2 m_\tau < M_a < m_\taup + m_\tau$, then the tree-level decay $a_\tau \to \tau^+ \tau^-$ has by far the 
largest branching fraction.
If the $\taup$ loop dominates, then the partial width for the decay $\atau \to \gamma\gamma$ is given by
\begin{equation}
\label{eq:width_atau_gammagamma}
\Gamma \left( \atau \to \gamma\gamma \right) = 
\frac{ \alpha^2  \, M_a^3 }{ 128\pi^3 \, m_\taup^2} \,   \left[ 
\left( {\rm Re} \, y_{\taup} \right)^2
+ \frac{4}{9} \left( {\rm Im} \, y_{\taup} \right)^2  \right]  \; ,
\end{equation}
to leading order in $M_a^2/(2  m_\taup)^2$.

Let us analyze the range of parameters where the  $\atau \to \gamma\gamma$ decay has the largest  branching fraction.
Clearly, this is true for $M_a < 2 m_\tau$. For larger $M_a$,
the ratio of partial widths can be written as 
\begin{equation}
\label{eq:width_atau_gammagamma_ratio}
\frac{ \Gamma \left( \atau \to \gamma\gamma \right) }{ \Gamma \left( a_\tau \to \tau^+ \tau^- \right)  } = 
\frac{ \alpha^2  \, M_a^2  \, \rho }{ 8\pi^2 \, m_\taup^2 \, s_L^2 } \,  
\end{equation}
where $\rho$ is a function of $y_o/y_\mathcal{E}$, $s_L m_{\tau'} /m_{\tau}$ and the phases $\phase_\mathcal{E}$ and $\phase_o$, given by 
\be
\rho = 
\frac{ \left(   \cos \phase_\mathcal{E}  + (y_o/y_\mathcal{E})  \,  s_L m_{\tau'} /m_{\tau}  \, \cos \phase_o \right)^2
+ \frac{4}{9} \left(  \sin \phase_\mathcal{E} + (y_o/y_\mathcal{E})  \, s_L m_{\tau'} /m_{\tau}  \, \sin \phase_o  \right)^2  }
{ \left| y_o / y_\mathcal{E}  - s_L m_{\tau'} /m_{\tau}  \, e^{i (\phase_\mathcal{E} - \phase_o) }  \right|^2 }    ~~.
\ee
Consider, for simplicity, the case where $ s_L \ll (m_{\tau} /m_{\tau'})  \, y_\mathcal{E}  / y_o $ and $ |\phase_\mathcal{E}| \ll 1 $, so that 
\be
\label{eq:width_atau_gammagamma_special}
\frac{ \Gamma \left( \atau \to \gamma\gamma \right) }{ \Gamma \left( a_\tau \to \tau^+ \tau^- \right)  }  \approx   \left( 9 \times 10^{-4} \, 
\frac{M_a  \, y_\mathcal{E} }{m_\taup  \, y_o \, s_L }  \right)^2  \,   ~~.
\ee 
Thus, if $s_L \lesssim 10^{-3}  (M_a /m_{\tau'})  \, y_\mathcal{E}  / y_o$, then $\atau \to \gamma\gamma$ is the dominant decay mode even for
$2 m_\tau < M_a \ll  2 m_\taup $.
In this case, $|y_{\taup}| \approx y_\mathcal{E}$ and   the proper decay length of $a_\tau$ is given by
\begin{equation}
\label{eq:ctau_benchmark}
c\tau_a \approx  4.0~\mathrm{cm} \, \times \, \left(\frac{0.1}{ 
|y_{\taup}|}\right)^{\! 2} \left(\frac{2~\mathrm{GeV}}{M_a}\right)^{\! 3} \, \left(\frac{m_\taup}{500~\mathrm{GeV}}\right)^{\! 2} \; .
\end{equation}
As we will see in the next Section, the above proper decay length may be large enough, when combined with the boost to the lab frame, 
to lead to $a_\tau$ decays a few meters away from the interaction point.

Tree-level mixing between $\varphi_\tau$ and the SM Higgs boson is induced by 
a  $\lambda |\phi|^2 H^\dagger H$ term in the scalar potential, where $H$ is the SM Higgs doublet and $\lambda$ is a dimensionless coupling.
Furthermore, a mixing between the $\varphi_\tau$ and $a_\tau$ particles is induced at one loop by their couplings (\ref{eq:pseudo})
to $\taup$ and $\tau$. Thus, besides the couplings to $\taup$ and $\tau$ discussed so far, the physical spin-0 particle that is mostly $a_\tau$ acquires the couplings of the SM Higgs boson times an overall factor $\epsilon$. 
As the measured properties of the observed Higgs boson of mass near 125 GeV fit well the SM predictions, a stringent upper limit on $\lambda$ can be derived. This, combined with the loop factor, implies $\epsilon  \ll 1$, and given the small $h^0$ couplings to light SM particles 
we conclude that the $a_\tau$-$h^0$ mixing can be ignored in what follows.

The $\varphi_\tau$-$h^0$ mixing, though, could be large enough (if $\lambda$ is not much below its upper limit) 
to lead to prompt $\varphi_\tau$ decays into $b\bar b$ and other final states associated with the SM Higgs boson. 
Nevertheless, it is more likely that the prompt decay $\varphi_\tau \to a_\tau a_\tau$ is the dominant one, due to 
the $|\phi|^4$ term in the scalar potential, which induces a sizable $\varphi_\tau a_\tau a_\tau$ coupling.

\bigskip

\section{Pseudoscalar signals at the LHC}
\label{sec:production}
\setcounter{equation}{0}

The LHC phenomenology of our model is dominated by the pair production $pp \to \taup^+ \taup^-$ of Vleptons which subsequently decay via $\taup \to \tau \atau,  \,  \nu W, \, \tau Z$, or $\tau h^0$. In this Section we first discuss various signal topologies that remain to be searched for at the LHC. Then we compute the production cross section, and the $ \atau$ decay length in the lab frame, in order to determine which detector systems are most suitable to search for the  $ \atau$ signals. 

\begin{figure}[t!]
	\begin{center}
		\hspace{-8mm} \includegraphics[width=0.94\textwidth]{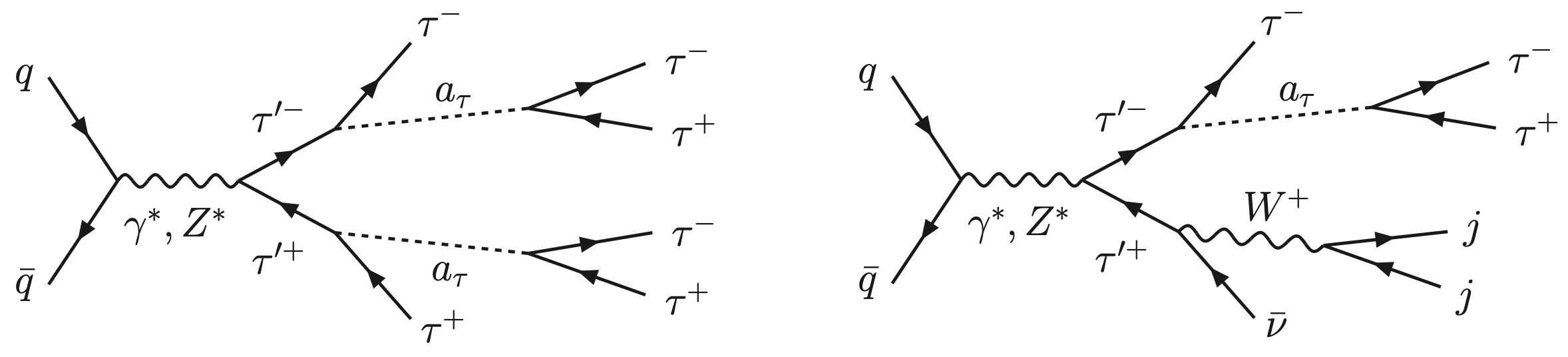} \\  [2mm]
		\hspace{8mm} \includegraphics[width=0.94\textwidth]{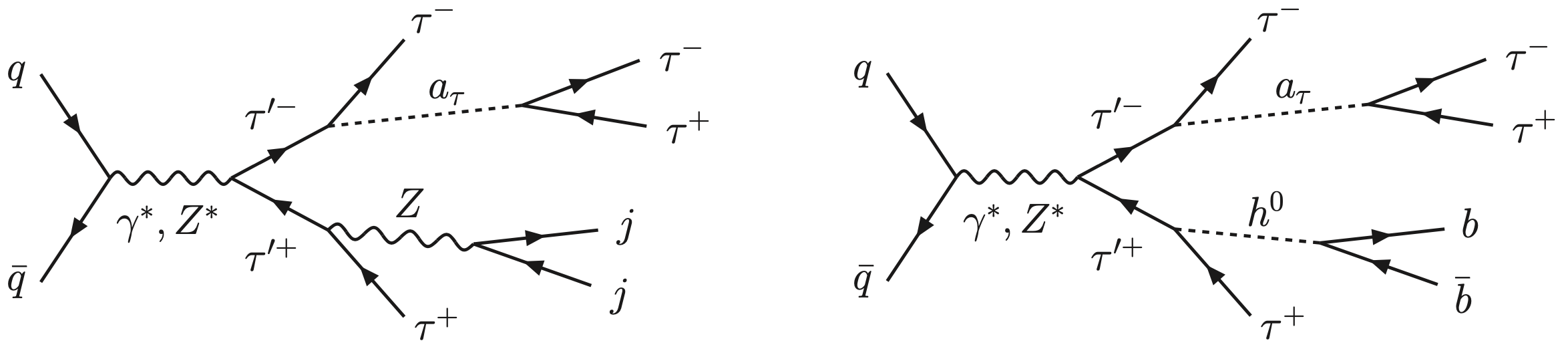}
		\caption{Vlepton pair production at the LHC leading to various final states: $6\tau$,  \ $3\tau + W \nu$, or $4\tau + Z/h^0$. Only processes involving at least one pseudoscalar $a_\tau$ are considered here, and only the dominant decay modes of the SM bosons are shown. Similar diagrams with one (or both) of the $a_\tau \to \tau^+\tau^-$ decays replaced by $a_\tau \to \gamma\gamma$ are important for the range of parameters where the $\gamma\gamma$ branching fraction is large [see Eq.~(\ref{eq:width_atau_gammagamma_ratio})].		 }
		\label{fig:diagramsMixed}
	\end{center}
\end{figure}

\begin{figure}[t!]
	\begin{center}
		\includegraphics[width=0.68\textwidth]{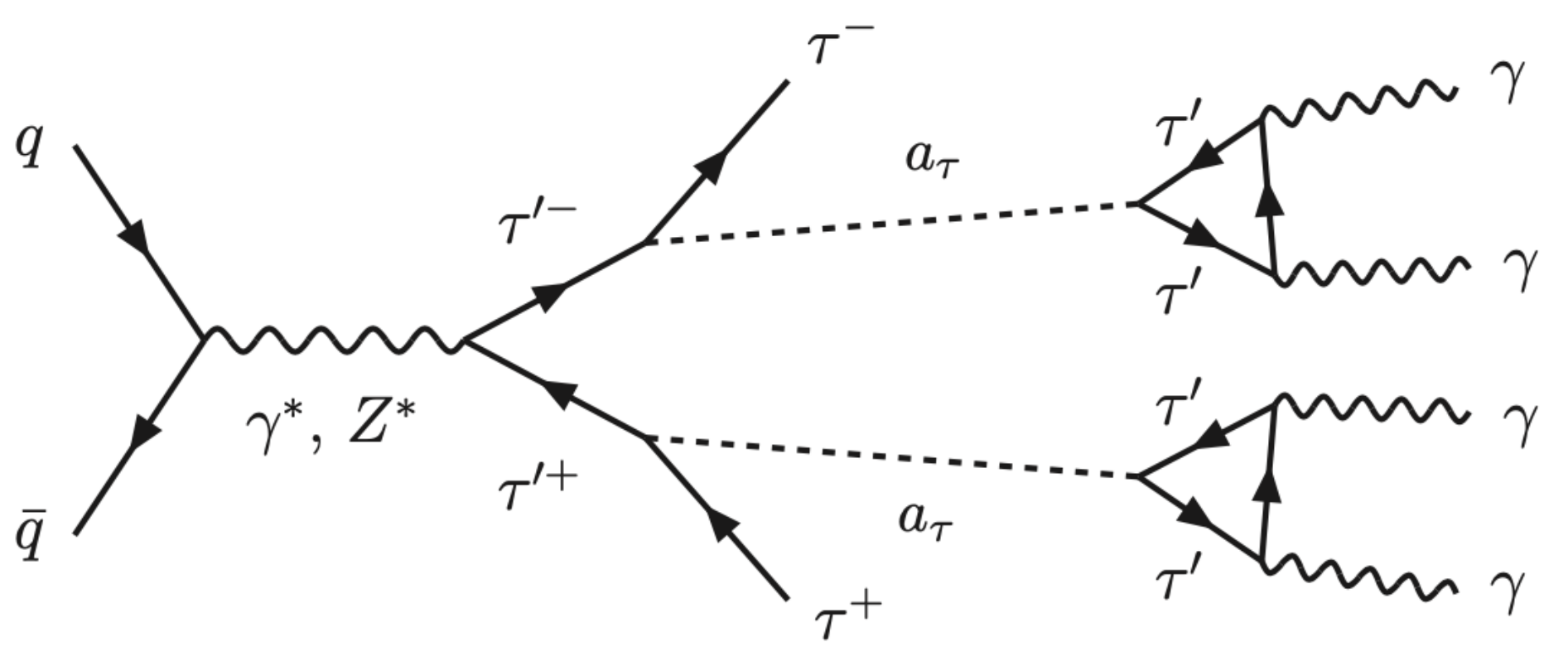}
		\caption{Electroweak production of Vleptons at the LHC, followed by cascade decays that give the $\tau^+\tau^-+ 4 \gamma$ final state analyzed in Section~\ref{sec:displaced}. The photons are displaced when the pseudoscalar $a_\tau$ is long lived (see Figure~\ref{fig:median_decaydist}). }
		\label{fig:diagram4gamma}
	\end{center}
\end{figure}

For the region of parameter space where the $B(\taup \to \tau \atau)$ branching fraction is the dominant one, {\it i.e.}, $R_\tau \gg 1$ where  $R_\tau$ is the ratio of widths defined in 
Eq.~(\ref{eq:Rtau}), the main process is $pp \to \taup^+ \taup^- \to \tau^+ a_\tau \tau^- a_\tau$. If $a_\tau$ decays predominantly to  $\tau^+ \tau^-$, the process of interest is shown in the first diagram of Figure~\ref{fig:diagramsMixed}, and leads to a striking $6\tau$ final state.
If the $a_\tau$ decay into photons has a large branching fraction, as analyzed in 
Eqs.~(\ref{eq:width_atau_gammagamma})-(\ref{eq:width_atau_gammagamma_special}), then the most relevant final state is 
$\tau^+\tau^-+ 4 \gamma$; this is shown in Figure~\ref{fig:diagram4gamma}, and is studied in detail in Section~\ref{sec:displaced}. 

If $B(\taup \to \tau \atau)$ is comparable to the $\taup \to \nu W$ branching fraction (which itself is approximately twice the 
branching fraction for $\tau Z$, or $\tau h^0$), {\it i.e.}, $R_\tau = O(1)$,  then it is preferable to consider the processes where only one 
of the Vleptons decays into $\tau \atau$, as shown in the last three diagrams of Figure~\ref{fig:diagramsMixed}. 
In these processes, the $\tau^+\tau^-$ pair arising from an $a_\tau$ decay may be displaced, depending on the proper decay length of $a_\tau$ [see Eq.~\ref{eq:ctau_benchmark})] and the $a_\tau$ boost, while the other final state particles are prompt.
For the range of parameters where the $a_\tau \to \gamma\gamma$  branching fraction is large, which follows from  Eq.~(\ref{eq:width_atau_gammagamma_ratio}) especially for the case of  lighter $a_\tau$,
the $\tau^+\tau^-$ pairs produced by $a_\tau$ in the diagrams of Figure~\ref{fig:diagramsMixed} are each replaced by a pair of nearly collinear photons.
The case where $B(\taup \to \tau \atau) \ll 1$ is covered by existing searches  \cite{CMS:2022nty}  for the standard decays of Vleptons.

If $2M_a < M_\varphi < m_\taup - m _\tau$, then the cascade decays of Vleptons into $\tau \varphi_\tau$ followed by 
$\varphi_\tau \to a_\tau a_\tau$ give rise to remarkable final states:     
\be
q \bar q \to \taup^+\taup^- \to (\tau^+ \varphi_\tau) (\tau^- \varphi_\tau) \to \tau^+\tau^- + 4 a_\tau   ~~,
\ee
where the $\tau$ pair is prompt, and the four $a_\tau$'s may have displaced decays into photons or $\tau$'s. Note that the most likely 
final states include $10\tau$ or $\tau^+\tau^-  + 8 \gamma$. 
The process in which one Vlepton decays into $\tau \varphi_\tau$ and the other one into $\tau a_\tau$ may also have a sizable rate,
and leads to $\tau^+\tau^-\! + 3 a_\tau$.

\smallskip 

\subsection{Vlepton production}
\label{sec:vlepton_prod}

Production of $SU(2)_W$-singlet Vleptons at the LHC proceeds mainly through an s-channel photon or $Z$ boson.
We simulate leading-order $\taup^+\taup^-$ production and the subsequent decay chain $\taup \to \tau a_\tau, a_\tau \to \gamma\gamma$ with \texttt{MadGraph5\_aMC{@}NLO}~2.6.4~\cite{Alwall:2014hca} using a UFO model generated with \texttt{FeynRules}~\cite{Alloul:2013bka} and the \texttt{NNPDF23\_lo\_as\_0130\_qed} PDF set~\cite{Ball:2013hta}. Decay positions for the long-lived pseudoscalar $a_\tau$ are generated in MadGraph using the decay width given in Eq.~\eqref{eq:width_atau_gammagamma}.  We use \texttt{Pythia~8}~\cite{Sjostrand:2014zea} to simulate showering and hadronization, which can lead to additional ISR jets modifying the $\taup$ production kinematics. Finally, we pass the generated events to \texttt{Delphes~3}~\cite{deFavereau:2013fsa} for detector simulation using its CMS card.

\begin{figure}[t!]
	\begin{center}
		\includegraphics[width=0.6\textwidth]{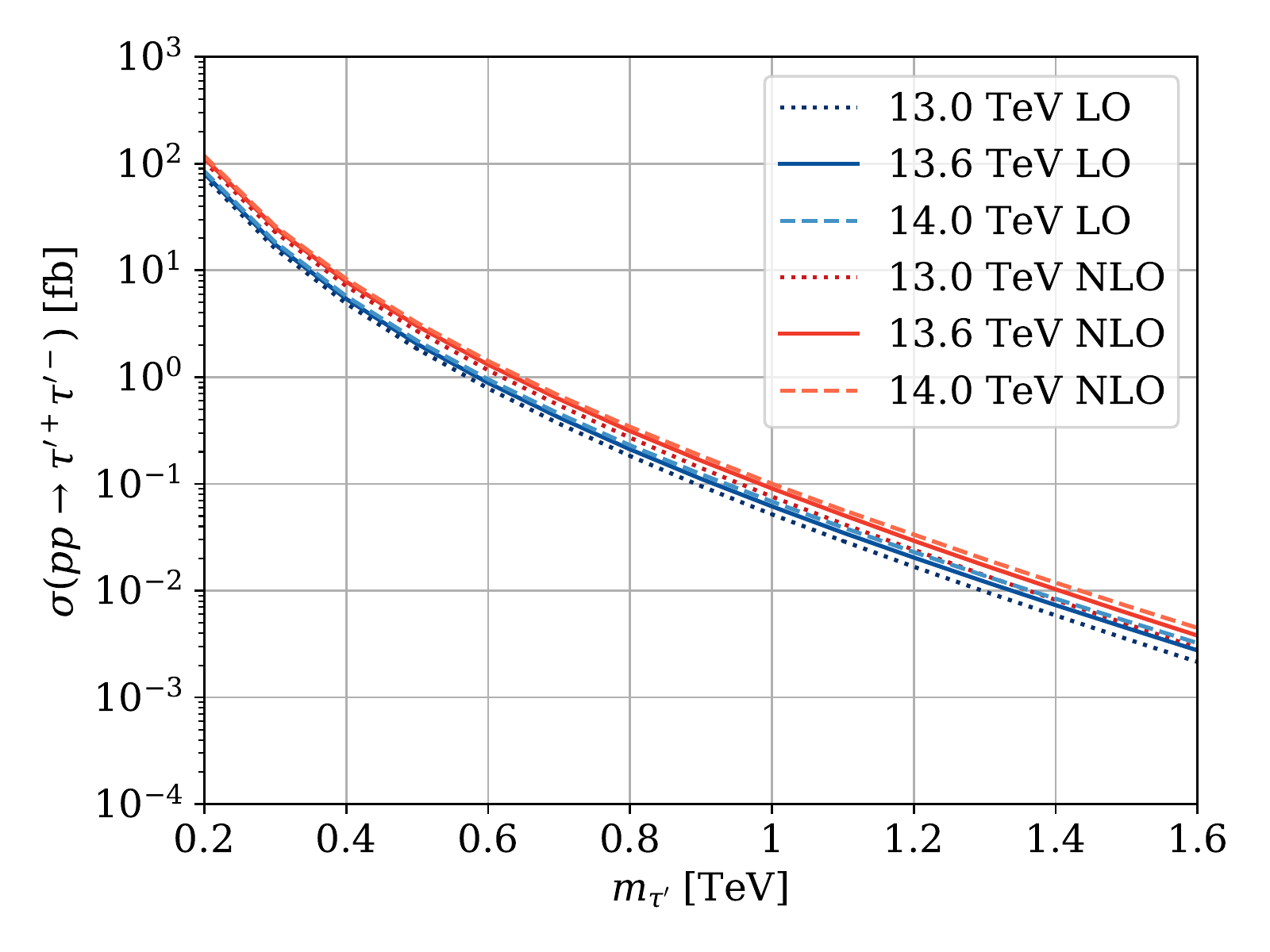}
		\caption{LHC cross section for pair production of weak-singlet Vleptons at different center-of-mass energies. NLO corrections increase the cross section by 25 to 50\%.}
		\label{fig:vll_singlet_xsec}
	\end{center}
\end{figure}

Since all couplings relevant for the $\taup^+\taup^-$ production process are fixed by SM gauge charges, the cross section depends only on the mass $m_\taup$ of the Vlepton. Figure~\ref{fig:vll_singlet_xsec} shows the pair production cross section at leading order for $SU(2)_W$ singlet Vleptons as a function of their mass $m_\taup$ for LHC center-of-mass energies of 13~TeV, 13.6~TeV and 14~TeV, which correspond, in turn, to the completed Run 2, the current Run 3, and the future high-luminosity runs. While the cross section drops rapidly with increasing mass, it can be larger than $10$~fb for $m_\taup \lesssim 300$~GeV, and larger than $1$~fb for $m_\taup \lesssim 600$~GeV.  Hence, hundreds or even thousands of events with final states shown in Figures~\ref{fig:diagramsMixed} or \ref{fig:diagram4gamma} could be produced by the end of the LHC Run 3.

To give an estimate of the size of next-to-leading order (NLO) corrections, Figure~\ref{fig:vll_singlet_xsec} also shows the NLO cross section for $\taup^+\taup^-$ production simulated with \texttt{MadGraph5\_aMC{@}NLO} version~3.4.2~\cite{Alwall:2014hca} using the UFO model of Ref.~\cite{Ajjath:2023ugn} and the \texttt{NNPDF23\_nlo\_as\_0118\_qed} PDF set~\cite{Ball:2013hta}. The renormalization and factorization scales are set to the invariant mass of the $\taup^+\taup^-$ pair. We find that NLO corrections increase the cross section by 25 to 50\% depending on the $\taup$ mass and the LHC energy.

\begin{figure}[t!]
	\begin{center}
		 \hspace*{-5mm} \includegraphics[width=0.525\textwidth]{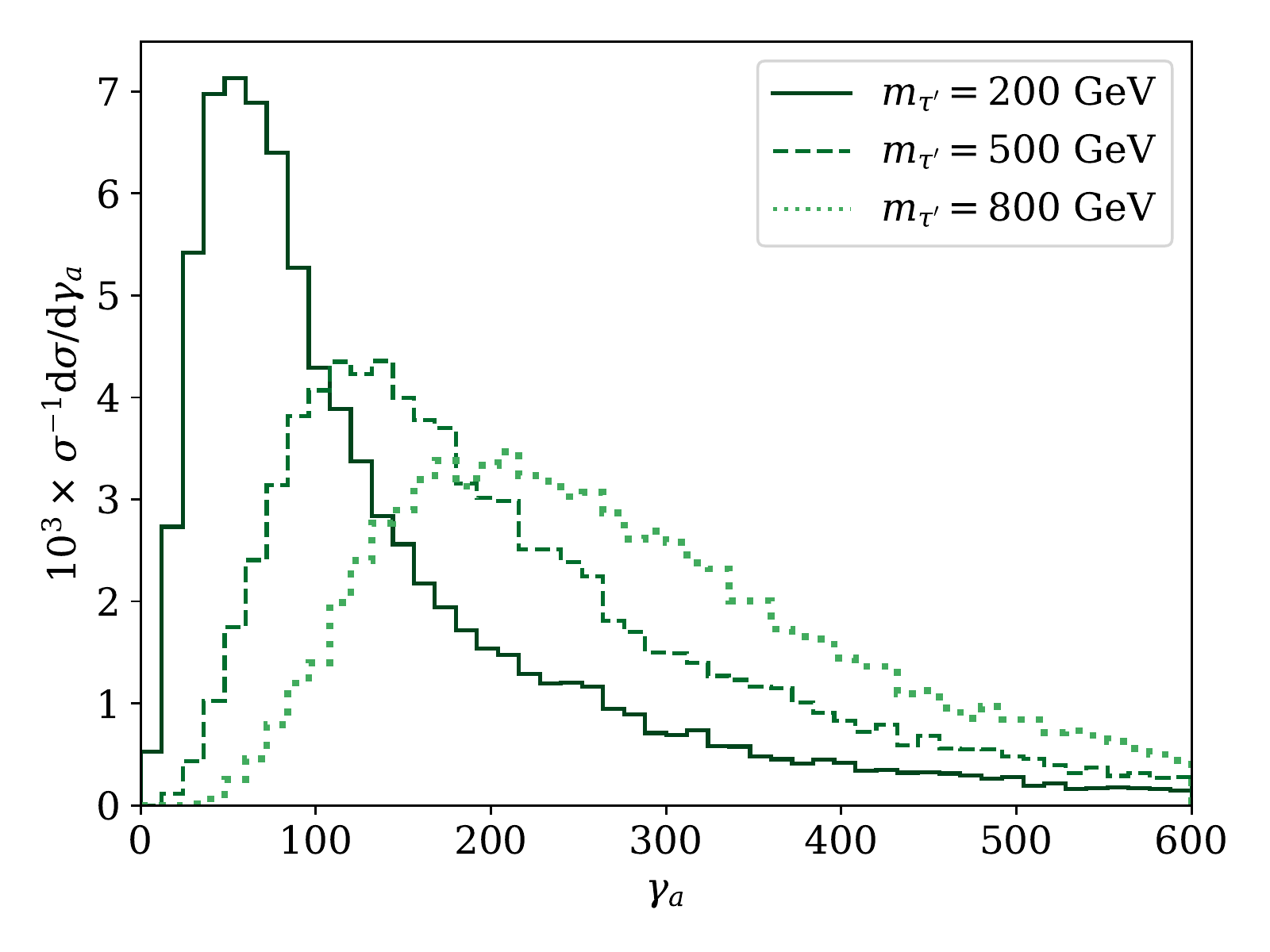}  \hspace*{-7mm}
		\includegraphics[width=0.525\textwidth]{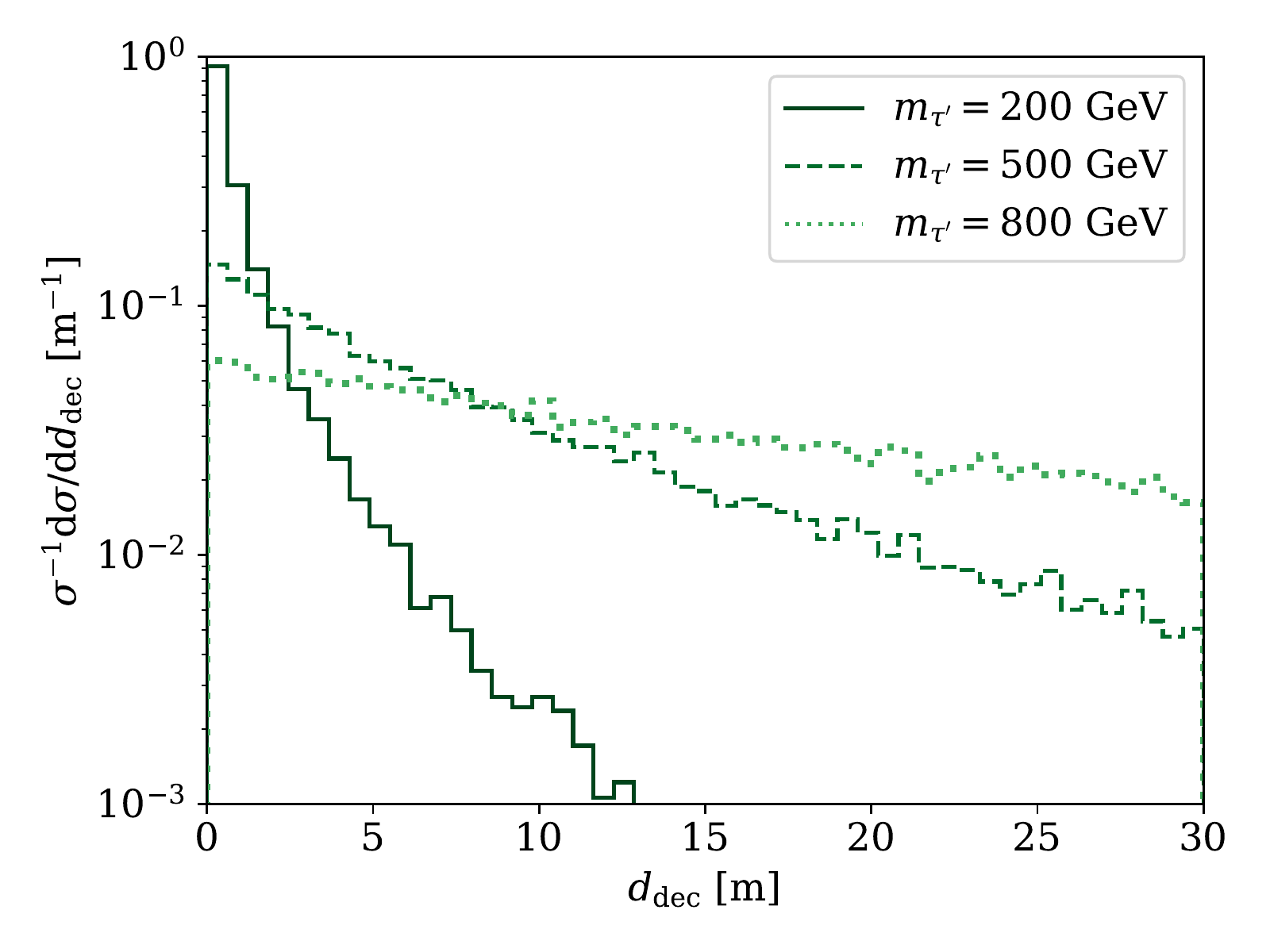}   \hspace*{-5mm}
		 \vspace*{-0.2cm} 
		\caption{Distributions of the boost (left panel) of $\atau$, and of the distance $d_{\rm dec}$ of $a_\tau$ decays from the interaction point (right panel) for $M_a=2$~GeV and different values of $m_\taup$ at the LHC at $\sqrt{s}=13.6$~TeV.
		For the $d_{\rm dec}$ distribution we assumed that $B(a_\tau \to  \gamma \gamma)$ is nearly 100\%, and that the Yukawa coupling present in Eq.~(\ref{eq:ctau_benchmark}) is $y_\taup=0.1$,  }
		\label{fig:llp_boost}
	\end{center}
\end{figure}

\subsection{Pseudoscalar decay length in the lab frame}

The fact that $a_\tau$ is produced via the decay of a heavy particle shapes its kinematic distributions. In particular, $a_\tau$'s produced this way are typically highly boosted.
The full distribution of boost factors $\gamma_a$ at the 13.6 TeV LHC is shown in the left panel of Figure~\ref{fig:llp_boost} for $M_a =2$~GeV and three values of $m_\taup$: 200~GeV, $500$~GeV, and $800$~GeV. Note that the average boost scales as $\langle \gamma_a\rangle \sim M_a^{-1}$.
The large pseudoscalar boosts that can be reached via this production mode result in a substantial enhancement of the decay length in the lab frame (shown in the right panel of Figure~\ref{fig:llp_boost}) over the proper decay length derived in Eq.~\eqref{eq:ctau_benchmark}.  Hence, we find that the majority of $a_\tau$ decays can easily occur several meters away from the interaction point. LLP decays in this distance range would be missed by searches for displaced decays in the tracker or calorimeters and would instead manifest themselves as activity in the muon chambers of CMS and ATLAS.

The muon system of CMS consists of cathode strip chambers (drift tubes) in the endcaps (barrel) alternating with layers of steel. This design makes it possible to use the CMS muon system as a sampling calorimeter, where electromagnetic or hadronic showers starting in the steel layers can be identified in the detector layers. In this way, CMS can have excellent sensitivity to both the $\tau^+ \tau^-$ and the $\gamma \gamma$ final state of $a_\tau$, with the efficiency being slightly larger for hadronic taus.

The muon system of ATLAS, on the other hand, contains less high-density material in which photons can convert to charged particles and start an  electromagnetic shower. Hence, the $\gamma \gamma$ final state may be more challenging to detect at ATLAS, while we expect sensitivity similar to CMS for decays into $\tau^+ \tau^-$. Thus, it would be useful if both collaborations performed searches for the final states shown in Figure~\ref{fig:diagramsMixed}.

\bigskip

\section{LHC sensitivity to highly displaced $a_\tau$ decays}
\setcounter{equation}{0}
\label{sec:displaced}

In this Section we study in detail the sensitivity of prospective and existing LHC searches to pseudoscalars produced in the decays of Vleptons, focusing on the process shown in Figure~\ref{fig:diagram4gamma}, which leads to a 
$\tau^+\tau^-+ 4 \gamma$ signal. That process is the most important one at the LHC when the branching fractions for 
$\tau^\prime \to \tau a_\tau$ and $a_\tau \to \gamma\gamma$ are large [see Eqs.~(\ref{eq:Rtau}) and (\ref{eq:width_atau_gammagamma_ratio})].

\begin{figure}[t!]
	\begin{center}
		 \vspace*{-0.5cm} 
		\includegraphics[width=0.6\textwidth]{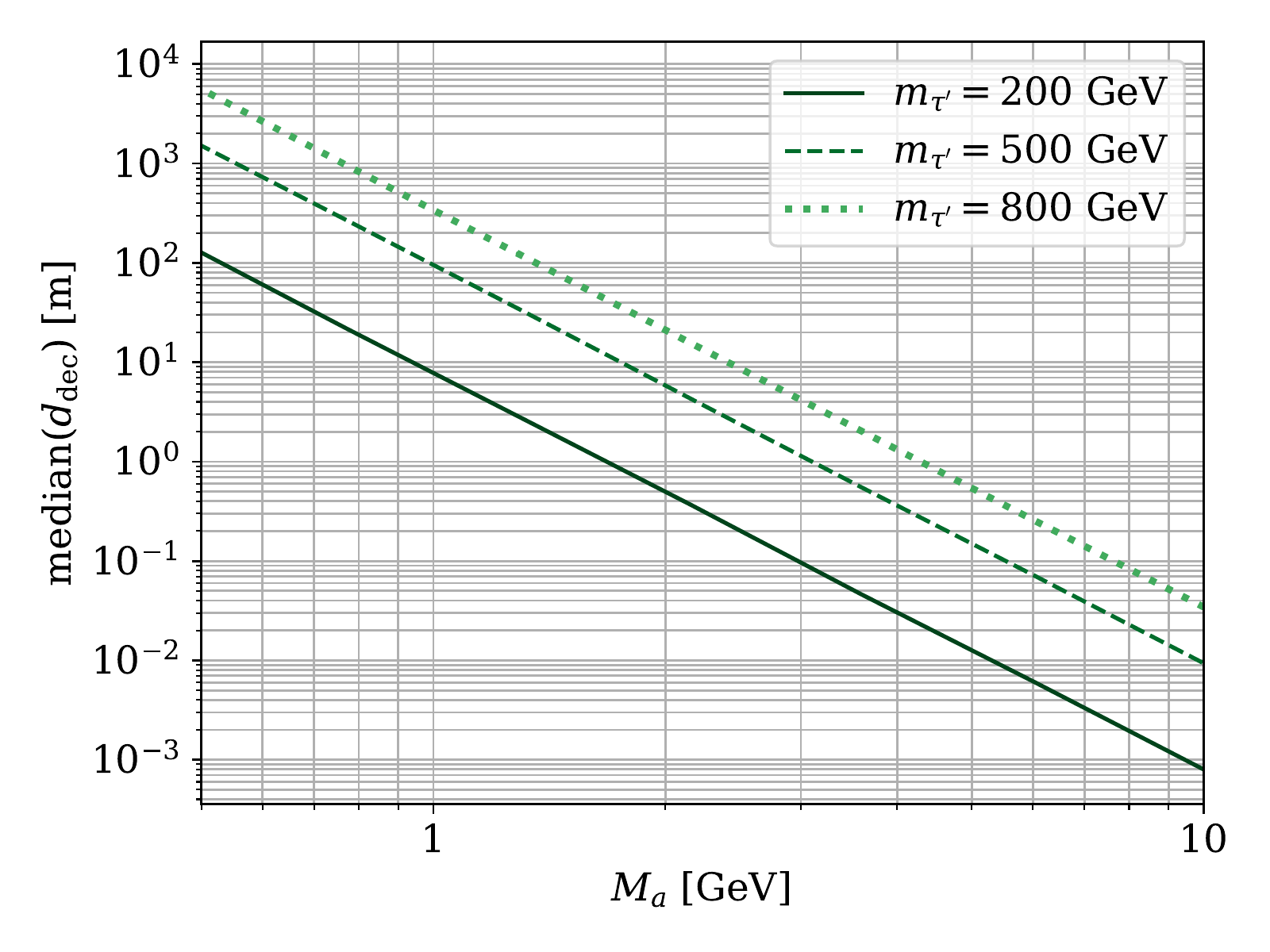}
		 \vspace*{-0.3cm} 
		\caption{Median decay length in the lab frame of an $a_\tau$ produced in the decay of a $\tau^\prime$ of mass 200, 500 or 800 GeV, at the 13.6 TeV  LHC. The decay length $d_{\rm dec}$ is computed here for $y_\taup=0.1$, and 
		changes for different Yukawa couplings as $|y_\taup|^{-2}$. \\ [-1cm] }
		\label{fig:median_decaydist}
	\end{center}
\end{figure}

The pseudoscalar $a_\tau$ has a sizable lifetime throughout large parts of parameter space [see Eq.~(\ref{eq:ctau_benchmark})] and is, additionally, produced with a large boost factor (see Section~\ref{sec:production}). Hence, as shown in Figure~\ref{fig:median_decaydist}, for $m_\taup \sim$ a few hundreds of GeV, the median decay length of a GeV-scale $a_\tau$ in the lab frame is several meters. Therefore, we mainly focus on searches for LLP decays in the muon system, which covers distances from the interaction point in the range of approximately 4~m to 12~m in CMS ~\cite{CMS:2008xjf }, and to 17~m in ATLAS \cite{ATLAS:2008xda}. 
Thus, the signal is two pairs of collimated photons produced in the muon system, and two prompt taus that can be used for triggering.
Complementary constraints from searches for decays inside the calorimeter and for missing energy, which are relevant for shorter and longer decay lengths, respectively, are discussed in Section~\ref{sec:existing_searches}.

\subsection{Prospective search for $a_\tau$ decays in the muon system}
\label{sec:muon_system_search}

Both ATLAS and CMS have recently carried out first searches for displaced decays in their muon chambers~\cite{ATLAS:2018tup, ATLAS:2019jcm, CMS:2021juv}. The CMS analysis presented in Ref.~\cite{CMS:2021juv} is particularly sensitive to light LLPs in a wide range of different final states (see, {\it e.g.}, \cite{Cottin:2022nwp}) as it searches for electromagnetic or hadronic cascades in the muon system instead of reconstructing displaced vertices.
In the following we propose a search for LLPs decaying in the muon system and discuss its sensitivity to long-lived pseudoscalars produced in Vlepton decays. Since we will base the muon system efficiencies and some selection criteria on the existing CMS search for displaced decays in the muon system, we begin by briefly summarizing those requirements.

At the trigger level, Ref.~\cite{CMS:2021juv} selects events if they have missing transverse energy $E_T^\mathrm{miss} > 120$~GeV. At the analysis level, this requirement is increased to $E_T^\mathrm{miss} > 200$~GeV. In addition, the analysis requires at least one jet with $p_T > 50$~GeV and $|\eta| < 2.4$. These selection cuts are motivated by the benchmark signal model considered by CMS, where LLPs $a$ originate in the exotic decays $h \to a a$ of the SM Higgs boson and, hence, highly energetic LLPs are typically only produced in association with an ISR jet.

LLP decays are detected if they lead to showers in the cathode strip chambers (CSCs) of the endcaps of the muon system, located at logitudinal distances $|z|$ between $6.7$~m and $11$~m and radial distances $r < 7~\text{m}$ from the interaction point and covering pseudorapidities $|\eta| < 2$. To be selected each event must have at least one reconstructed shower at an azimuthal angle distance $\Delta \phi(\text{shower}, \vec{p}_T^\mathrm{miss}) < 0.75$ from the missing momentum vector. This requirement is motivated by the fact that particles decaying outside the calorimeter appear as missing energy even if they go on to decay in the muon detector. In addition, the shower must lead to at least $130$ hits in the CSCs. For recasting purposes, this requirement is accounted for in efficiencies, parametrized in terms of the energy of the LLP, provided by CMS~\cite{hepdata.104408}.

The existing CMS search for clusters from LLP decays in the muon system features a number of analysis choices that are motivated by the benchmark signal of Ref.~\cite{CMS:2021juv} and are suboptimal for other LLP models with comparable lifetimes.  In particular, the Vlepton model studied in this work differs from the CMS benchmark in a couple of essential aspects. In particular, the $a_\tau$ particles are produced in association with taus, which can be leveraged for triggering. Thus, the trigger requirements can be satisfied even in the absence of an additional boost caused by an ISR jet.

In the following we explore the efficiencies of different trigger configurations for our signal.
Since we are interested in the regime where the $a_\tau$ decays beyond the calorimeter, the LLP itself is counted as missing energy even if it goes on to decay visibly in the muon system. Hence, the final state objects available for triggering are missing transverse energy and -- depending on the decay products of the $\tau$ leptons in the event -- hadronic taus or leptons.

\begin{figure}[t!]
	\begin{center}
		\includegraphics[width=0.65\textwidth]{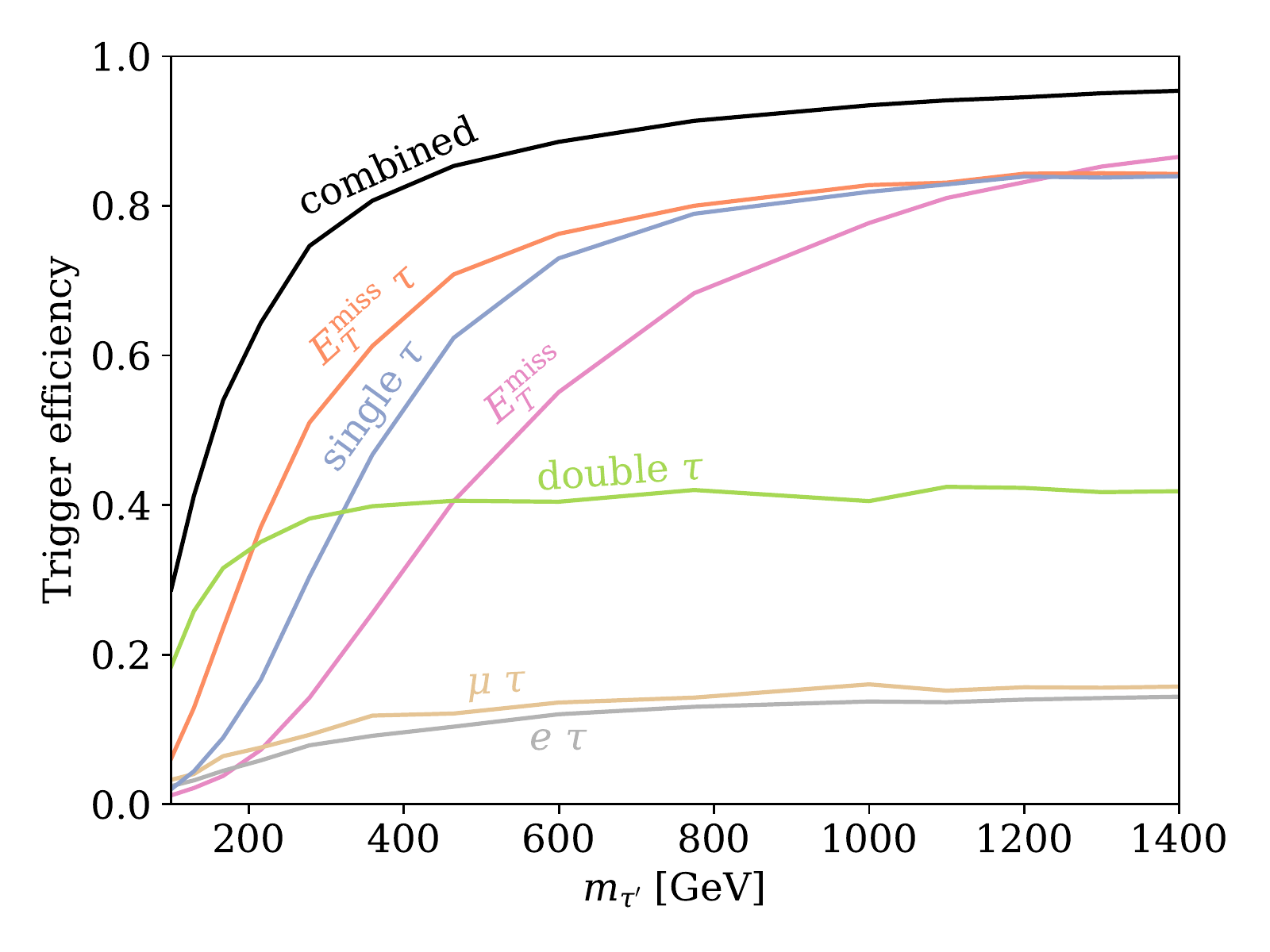}
		\caption{Efficiencies of the triggers listed in Table~\ref{tab:triggers} for events (at the 13.6 TeV LHC) where both $\atau$'s decay beyond the electromagnetic calorimeter, and thus are treated as missing energy. 
		The black line labelled ``combined" represents the efficiency obtained by requiring at least one of the trigger criteria to be fulfilled.}
		\label{fig:trigger_efficiencies}
	\end{center}
\end{figure}

Possible trigger conditions at CMS \cite{triggers} using different combinations of these final state objects are summarized in Table~\ref{tab:triggers}. Efficiencies of these triggers for signal events where both $\atau$'s decay outside the electromagnetic calorimeter are shown in Figure~\ref{fig:trigger_efficiencies} assuming full efficiency for hadronic $\tau$ identification. All triggers considered here become more efficient for larger $m_\taup$. However, the efficiency of the \emph{$E_T^\mathrm{miss}$} trigger continues to grow while the triggers involving taus approach plateaus set by the hadronic or leptonic branching fractions of the $\tau$ lepton. For $m_\taup \lesssim 200$~GeV, we find that the \emph{double $\tau$} trigger has the highest efficiency, while for larger masses the \emph{$E_T^\mathrm{miss}$ $\tau$} trigger is more efficient. For even larger masses, $m_\taup \gtrsim 1.2$~TeV, the latter is eventually overtaken by the pure \emph{$E_T^\mathrm{miss}$} trigger, which is not limited by $\tau$ branching fractions. Nevertheless, for most of the mass range that we are interested in the  \emph{$E_T^\mathrm{miss}$ $\tau$} configuration is the most efficient and dominates the total trigger efficiency that is obtained by requiring at least one of the different trigger conditions to be satisfied. This total efficiency is labeled as \emph{any trigger} in Figure~\ref{fig:trigger_efficiencies}.

\begin{table}[t!]
	\begin{center}
		\begin{tabular}{rc}
			\toprule
			\multicolumn{1}{c}{Trigger}  &  Condition \\
			\midrule
			$E_T^\mathrm{miss}$ & $E_T^\mathrm{miss} > 200$~GeV \\
			$e$ $\tau_\mathrm{h}$ & one $e$ with $p_T > 24$~GeV and one $\tau_\mathrm{h}$ with $p_T > 30$~GeV, $|\eta| < 2.1$ \\
			$\mu$ $\tau_\mathrm{h}$ & one $\mu$ with $p_T > 20$~GeV, $|\eta|<2.1$ and one $\tau_\mathrm{h}$ with $p_T > 27$~GeV, $|\eta| < 2.1$ \\
			$E_T^\mathrm{miss}$ $\tau_\mathrm{h}$ & $E_T^\mathrm{miss} > 100$~GeV and one $\tau_\mathrm{h}$ with $p_T > 50$~GeV \\
			single $\tau_\mathrm{h}$ & one $\tau_\mathrm{h}$ with $p_T > 180$~GeV and $|\eta| < 2.1$ \\
			double $\tau_\mathrm{h}$ & two $\tau_\mathrm{h}$'s with $p_T > 40$~GeV \\
			\bottomrule
		\end{tabular}
	\end{center}
	\caption{\label{tab:triggers} Possible trigger configurations for our displaced 
	 signal arising from the $\tau^+ a_\tau \tau^- a_\tau$ intermediate state. 
	 	 Hadronically decaying $\tau$'s are labelled $\tau_\mathrm{h}$.}
\end{table}

Another aspect in which our model substantially differs from the CMS benchmark is the centrality of the LLP decay positions in the detector. The current CMS analysis only extends to the endcaps of the muon system and excludes its barrel region. This choice still leads to reasonably good geometric acceptance if the LLP signal is strongly peaked in the forward direction. However, the $a_\tau$ is produced in the decays of Vleptons with masses of several hundreds of GeV, which leads to a rapidity distribution that is much more centrally peaked than that of LLPs from Higgs decays.

\begin{figure}[t!]
	\begin{center}
		 \hspace*{-1.7cm}  \includegraphics[width=0.525\textwidth]{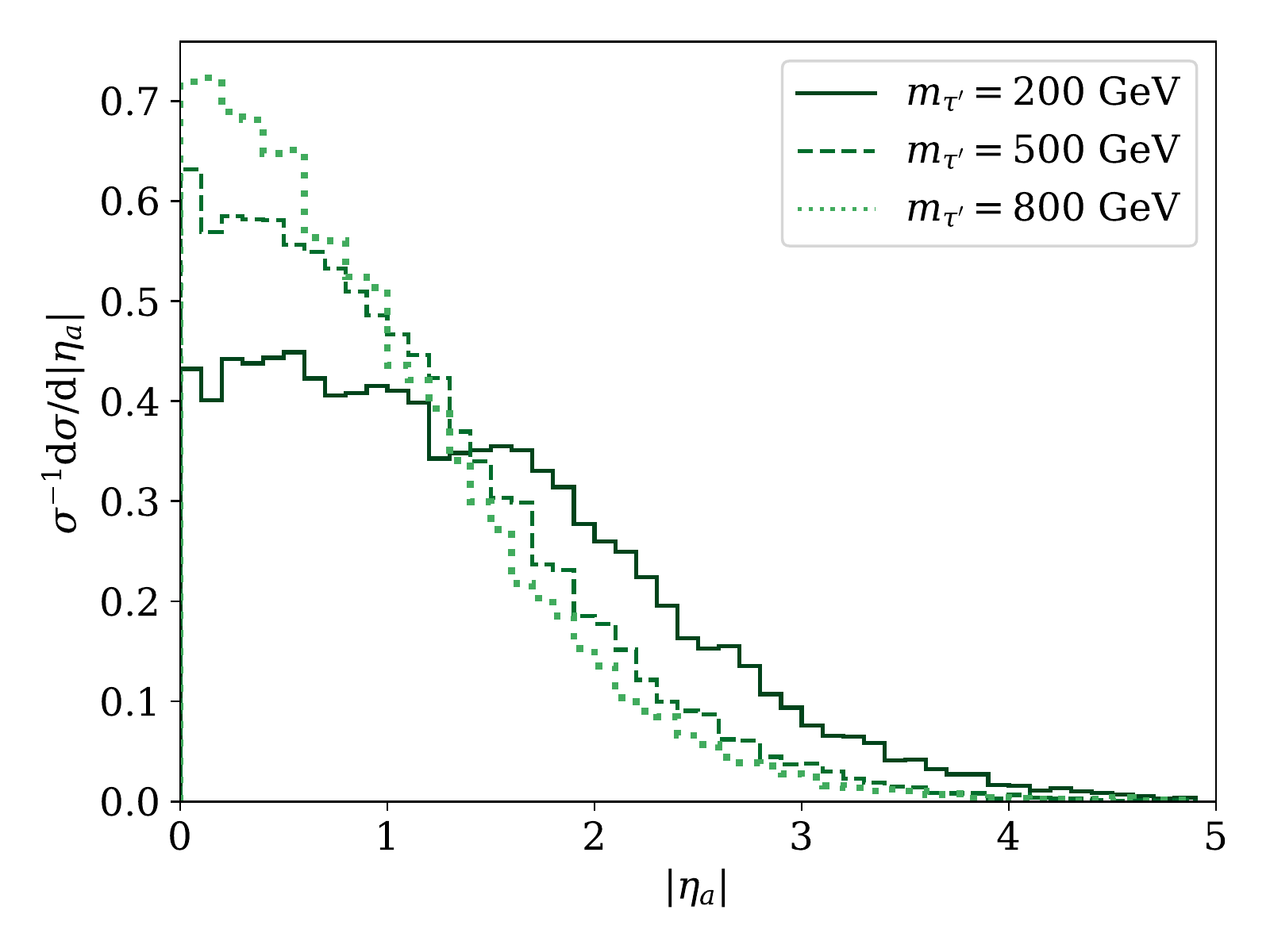}   \hspace*{-.6cm} 
		\includegraphics[width=0.525\textwidth]{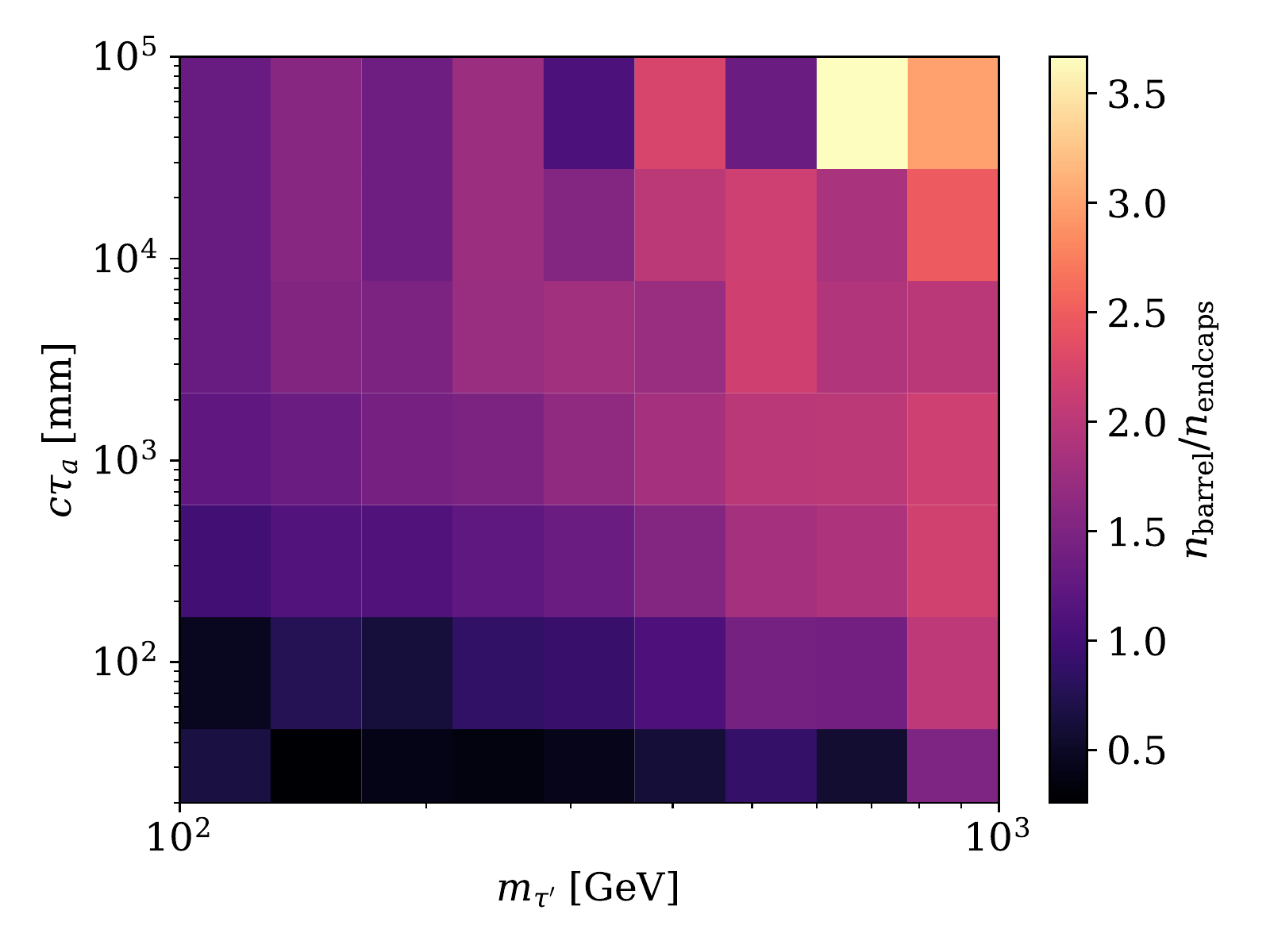}  \hspace*{-1.7cm} 
		\caption{Left panel: $a_\tau$ pseudorapidity distribution for different Vlepton masses $m_\taup$ and fixed pseudoscalar mass $M_a = 2$~GeV. Right panel: Ratio of events with at least one LLP decay within the geometric acceptance of the muon system barrel to events with at least one decay in the endcaps (shown for $y_\taup=0.01$, with the value of $M_a$ at each point following from Eq.~\eqref{eq:ctau_benchmark}).}
		\label{fig:barrel_vs_endcap}
	\end{center}
\end{figure}

The pseudorapidity distribution for pseudoscalar LLPs in our model is shown in the left panel of Figure~\ref{fig:barrel_vs_endcap}. Clearly, the distribution becomes more central for larger $m_\taup$, which leads to more LLPs decaying in the barrel region of the muon system rather than its endcaps. This is reflected in the right panel of Figure~\ref{fig:barrel_vs_endcap}, which shows the ratio of events with at least one decay within the geometric acceptance of the barrel to events with at least one decay within the geometric acceptance of the endcaps. We take the sensitive region of the barrel to be at $|z| < 6.6$~m and $4.6~\text{m} < r < 7.4~\text{m}$. As expected, the benefit of including decays in the barrel is greater for larger $m_\taup$. In addition, we find that it grows with increasing proper LLP decay length $c\tau_a$. This makes it imperative to also include the barrel region of the muon system alongside the endcaps in the analysis.

We are now in a position to investigate the sensitivity of a prospective search for LLPs that decay in the muon system and are produced in association with tau leptons. Following the discussion above, we require events to pass at least one of the triggers listed in Table~\ref{tab:triggers}. Moreover, we require at least one LLP decay per event in either the endcaps or the barrel of the muon system. For our projection, we approximate the cluster reconstruction efficiency in the barrel as identical to the endcaps. Otherwise we assume the same selection criteria as the existing CMS search, with the exception of the original $E_T^\mathrm{miss}$ requirement, which is superseded by trigger requirements from Table~\ref{tab:triggers}.

\begin{figure}[t!]
	\begin{center}
		\includegraphics[width=0.7\textwidth]{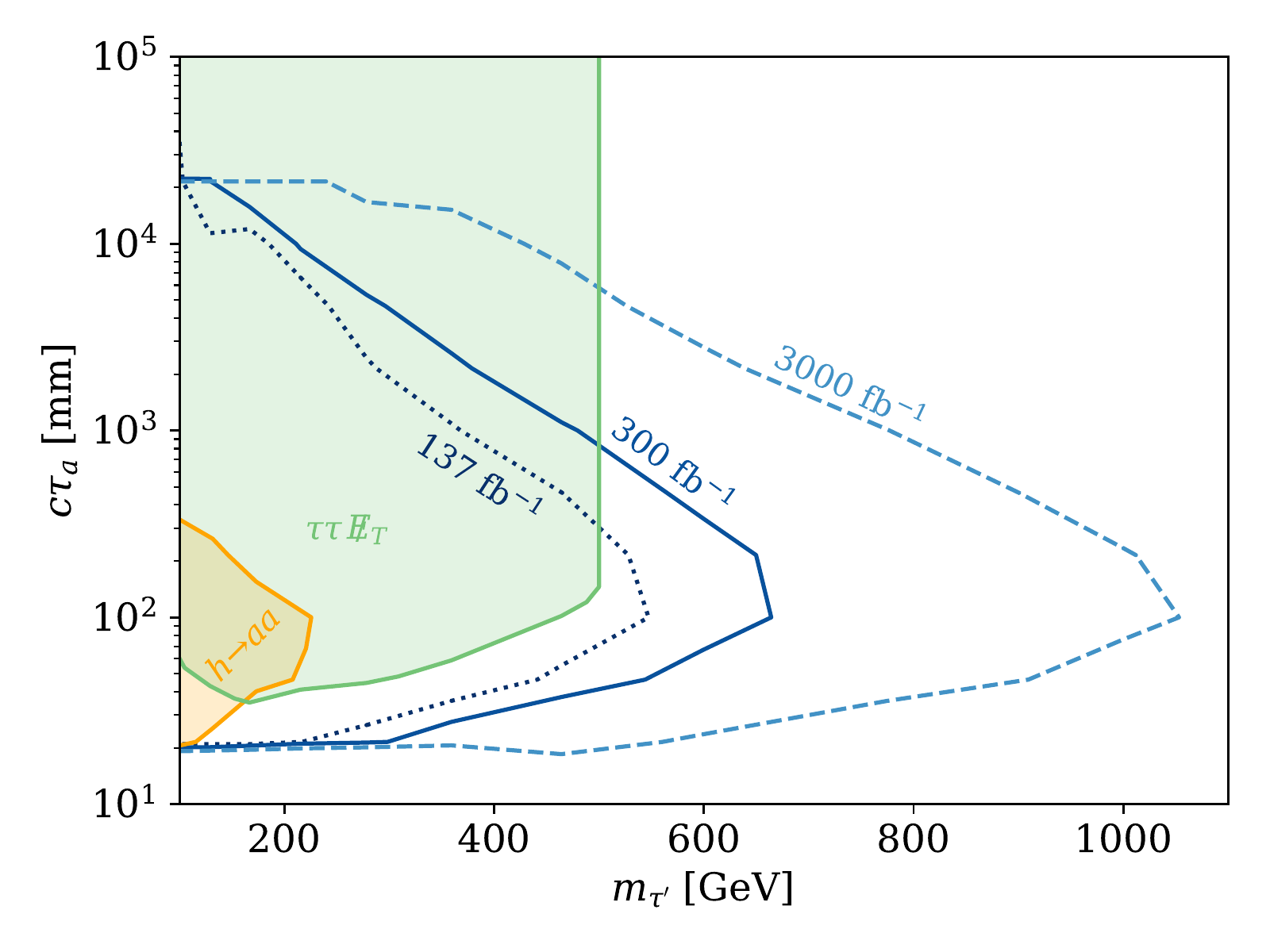}
		\caption{Projected sensitivity of a prospective search for displaced $a_\tau \to \gamma\gamma$ decays in the muon system in association with prompt taus (blue lines). Shaded regions are ruled out by 
				our recasts of existing CMS searches for displaced decays in the muon system~\cite{CMS:2021juv}  (orange), and for tau pair production in association with missing energy~\cite{CMS:2021woq}  (green). All constraints and projections are shown for $y_\taup=0.01$ (with the value of $M_a$ at each point following from Eq.~\eqref{eq:ctau_benchmark}).}
		\label{fig:limits_muonsystem}
	\end{center}
\end{figure}

\begin{figure}[t!]
	\begin{center}
		\includegraphics[width=0.7\textwidth]{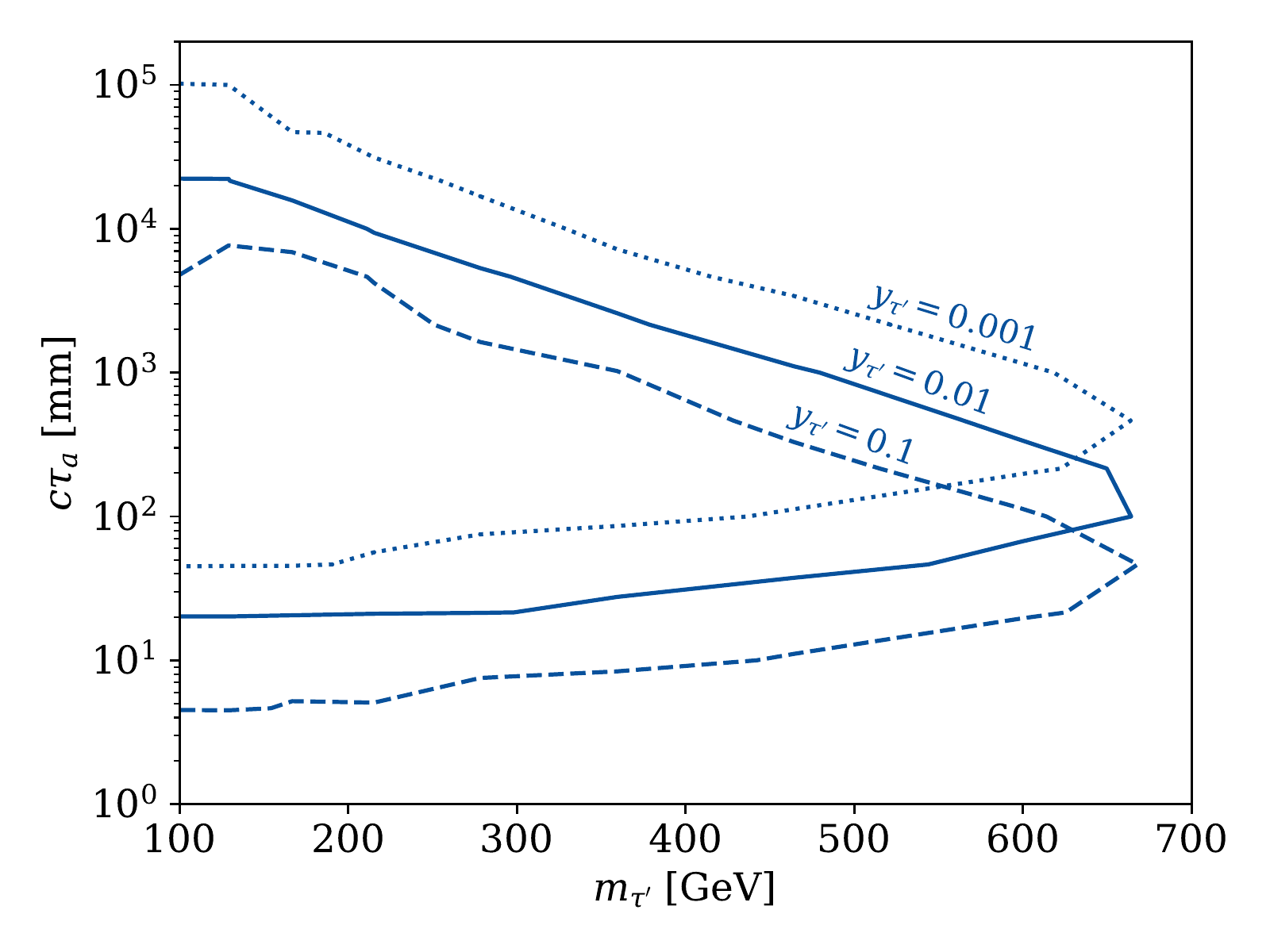}
		\caption{Projected sensitivities (corresponding to $N_\mathrm{sig}=3$) of a prospective search for displaced $a_\tau$ decays in the muon system in association with prompt taus at the 13.6 TeV LHC with $300$~fb$^{-1}$ of data, for different values of $y_\taup$ (with the value of $M_a$ at each point following from Eq.~\eqref{eq:ctau_benchmark}). The search is sensitive at the 95\% CL to each region to the left of the corresponding line.}
		\label{fig:limits_muonsystem_ytaup}
	\end{center}
\end{figure}

With these selection requirements we expect backgrounds to be negligible, at least before the HL-LHC run. Hence, assuming that no events in the signal region are observed, the projected 95\% CL exclusion extends to signal parameter points that predict an expected number of signal events $N_\mathrm{sig} > 3$. The resulting projected sensitivity is shown in Figure~\ref{fig:limits_muonsystem} for integrated luminosities of $137$~fb$^{-1}$ (corresponding to LHC run 2) and $300$~fb$^{-1}$ (corresponding to the end of LHC run 3) for center-of-mass energies of 13~TeV and 13.6~TeV, respectively. In addition, we show a projection for the HL-LHC run with $3$~ab$^{-1}$ and a center-of-mass energy of 14~TeV. Since it is not clear if the negligible-background assumption will be accurate at $3$~ab$^{-1}$, the corresponding projection in Figure~\ref{fig:limits_muonsystem} constitutes only an upper bound on the realistic expected sensitivity. These sensitivities as well as all projections and constraints discussed in the following are calculated based on Monte Carlo events simulated at leading order as described in Sec.~\ref{sec:vlepton_prod}.

The search is sensitive to Vlepton masses above 500~GeV already at a luminosity of 137~fb$^{-1}$. Moreover, it covers several orders of magnitude in the LLP proper decay length, from approximately 20~mm to more than 10~m. Note that the LLP is typically produced with substantial boost (see also Figure~\ref{fig:llp_boost}) and, hence, proper decay lengths of e.g.\ 20~mm generally correspond to much larger decay lengths in the lab frame. For higher integrated luminosities the sensitivity is improved further, especially to longer decay lengths and larger masses. Thus, the reach of the search extends up to nearly $m_\taup=700$~GeV at 300~fb$^{-1}$ and to approximately 1.1~TeV at 3~ab$^{-1}$.

In Figure~\ref{fig:limits_muonsystem} we set $y_\taup=0.01$. Projections for different values of $y_\taup$ (at $300$~fb$^{-1}$ and $\sqrt{s}=13.6$~TeV) are shown in Figure~\ref{fig:limits_muonsystem_ytaup}. As $y_\taup$ changes, the sensitivity projections shift with respect to $c\tau_a$ as the corresponding mass, and thus typical boost, of $a_\tau$ changes. However, the reach in $m_\taup$ remains unaffected.

For comparison, Figure~\ref{fig:limits_muonsystem} also shows the constrained derived from a recast of the existing CMS search~\cite{CMS:2021juv} for displaced decays in the muon system with an integrated luminosity of $137$~fb$^{-1}$. The selection requirements are summarized at the beginning of Section~\ref{sec:muon_system_search}.
In contrast to our expectation for the prospective search proposed in this Section, the existing search is not free of background. The dominant background sources are punch-through jets, muon bremsstrahlung and long-lived SM hadrons. With the selection cuts described above, $2 \pm 1$ background events in the signal region were expected and $3$ events were observed. To derive an approximate exclusion limit on the number of signal events, we construct the Poisson likelihood
\begin{equation}
	L = \frac{\lambda(\mu, \theta_B)^N}{N!} \, e^{-\lambda(\mu, \theta_B)} \, e^{-\theta_B^2/2} \; .
\end{equation}
Here $N$ denotes the number of observed events in the signal region, $\theta_B$ is a Gaussian nuisance parameter accounting for the background uncertainty, and $\lambda(\mu, \theta_B) = \mu S + B (1 + (\Delta B/B) \,  \theta_B)$, where $B$ is the expected number of background events and $S$ the expected number of signal events for a given signal hypothesis. Using pseudoexperiments we find that signal hypotheses predicting $S > 6.1$ signal events are excluded at 95~\% CL, which is the bound we use for the recast of the search shown in Figure~\ref{fig:limits_muonsystem}.

\subsection{Complementary constraints from existing searches}
\label{sec:existing_searches}

While searches in the muon system have unique sensitivity to LLPs that decay several meters away from the interaction point, there are several existing LHC searches that are sensitive to longer or shorter decay lengths. In the following we discuss constraints on our model of pseudoscalars from Vleptons based on the most relevant such searches.

\subsubsection*{Longer $a_\tau$ lifetime}

If the $a_\tau$ has a very long lifetime, most of its decays occur outside the LHC detectors and only manifest themselves as missing energy.
Additionally, since the visible energy in a collision is calculated based only on the momenta of particles detected in the tracker or calorimeter, LLPs that decay in the muon system are also still counted as missing energy despite becoming visible inside the detector. Hence, any signature consisting of displaced decays in the muon system $+X$ is also at the same time a $E_T^\mathrm{miss} + X$ signature. For the model discussed in this work the associated missing energy signature is $E_T^\mathrm{miss} + \tau \tau$.

The currently strongest constraint on this final state is provided by the CMS search for supersymmetric stau pair production, followed by the decay $\tilde{\tau} \to \tau \tilde{\chi}^0$ to taus and invisible neutralinos, in 137~fb$^{-1}$ of data~\cite{CMS:2021woq}.

This signature looks the same as events in our model where both LLPs decay after the electromagnetic calorimeter. The kinematics are also identical, besides small differences due to the different spins of the decaying particles and their decay products. We can use this observation to derive a conservative constraint on our model based on the limits from the stau search without fully recasting it. To this end, we estimate that a parameter point in our model is excluded if
\begin{equation}
	\sigma_{\taup}^\mathrm{prod} f_\mathrm{inv} \geq \sigma_{\tilde{\tau}_R}^\mathrm{lim} \; .
\end{equation}
Here, $\sigma_{\taup}^\mathrm{prod}$ is the $\taup$ production cross section, $f_\mathrm{inv}$ is the fraction of events in our model where both LLPs decay after the electromagnetic calorimeter, and $\sigma_{\tilde{\tau}_R}^\mathrm{lim}$ is the limit on the production cross section of right-handed staus set by Ref.~\cite{CMS:2021woq}. Note that this way of deriving a constraint is conservative, since a small fraction of events where only one LLP decays after the calorimeter may also fulfill the selection criteria of the stau search. Here we neglect this contribution.

The result of our reinterpretation of the stau search in terms of our model is shown in green in Figure~\ref{fig:limits_muonsystem} on top of the constraints and projections discussed in Sec.~\ref{sec:muon_system_search}.
Note that the constraint ends at $m_\taup = 500$~GeV since this is the largest mass for which Ref.~\cite{CMS:2021woq} provides a limit.
Comparing the reinterpreted limits from the stau search to our projections for a prospective LLP search in the muon system, we find that the stau search is more sensitive to long LLP decay lengths. This is of course obvious for very long decay lengths, where both LLP decays typically happen outside of the detector (including the muon system). The LLP search, in contrast, has greater reach to shorter decay lengths. Importantly, the LLP search also has far lower background than the corresponding missing energy search and is therefore expected to scale more favorably to higher luminosities.

\subsubsection*{Shorter $a_\tau$ lifetime}

For sufficiently short lifetimes of the $a_\tau$, it typically decays inside the calorimeter. There are a number of existing ATLAS and CMS searches for photon signatures that become relevant in this regime. LLPs decaying into photons within the radius of the electromagnetic calorimeter can give rise to a delayed photon signature. CMS has carried out a search for this signature in association with at least three jets in Ref.~\cite{CMS:2019zxa}. The search is sensitive to delayed photons that are detected in the calorimeter with a delay $\Delta t \gtrsim 1.5$~ns with respect to a particle produced at the interaction point and moving at the speed of light. In our model, photons produced in $a_\tau$ decays are rarely delayed by this much since the $a_\tau$'s are typically highly boosted. In particular, for distances $\lesssim 3$~m, a delay of $\Delta t \gtrsim 1.5$~ns requires $\gamma_a \lesssim 2$, which is only the case in a negligibly small fraction of events in our model (see Figure~\ref{fig:llp_boost}). Therefore, we do not consider this search here.

Due to the large typical boost factors our model also only very rarely produces a signal with three or more separated photons in the calorimeter, which is a signature that ATLAS searched for at $\sqrt{s}=8$~TeV with 20.3~fb$^{-1}$ of data~\cite{ATLAS:2015rsn}. Concretely, to be resolved separately the three photons have to be pairwise separated by at least $\Delta R_{\gamma\gamma} \gtrsim 0.3$. Since $\Delta R_{\gamma\gamma} \sim 2/\gamma_a$, where $\gamma_a$ denotes the boost of the decaying $a_\tau$, this requires $\gamma_a \lesssim 6$, which is only the case in a tiny fraction of events in the mass ranges of $\taup$ and $a_\tau$ that we focus on here. For this reason, in addition to the lower c.o.m.\ energy and luminosity, we do not expect this search for three photons to set a competitive bound on our model.

A highly boosted $a_\tau$ decaying into photons with $\Delta R_{\gamma\gamma} \lesssim 0.01$ would instead manifest itself as a collimated photon jet. ATLAS has carried out a search for this signature at $\sqrt{s} = 13$~TeV in $36.7$~fb$^{-1}$ of data~\cite{ATLAS:2018dfo}. However, this analysis targets a heavy resonance decaying into two photon jets and therefore searches for a bump in the spectrum of the di-photon-jet mass $m_{\gamma_R\gamma_R}$. Since our signature of two Vleptons each decaying into a pair of collimated photons and a tau does not produce such a bump, this search is not sensitive to our model. However, we emphasize that a search for collimated photons in association with tau leptons could have excellent sensitivity to this model and complement the muon system search discussed in this work in the parameter space with shorter $a_\tau$ decay lengths.

\bigskip

\section{Conclusions}
\setcounter{equation}{0}
\label{sec:Conclusions}

Vectorlike fermions are a major target for LHC searches, and the data sets accumulated by the ATLAS and CMS experiments have recently become large enough to allow the exploration of interesting mass ranges even for particles produced solely via electroweak interactions, like the Vlepton discussed here.
However, the decay modes of such fermions are model dependent, and may lead to final states that could escape generic searches.
It is thus imperative for theoretical studies to cover as many decay modes as possible, and for the experimental collaborations to perform dedicated searches for each of these possibilities.

The model studied in this paper is particularly simple, as it includes a single complex scalar besides the Vlepton and the SM. These particles may be motivated by models of compositeness, larger gauge symmetries, or other theoretical constructions, but from a phenomenological point of view they may be treated independently of their origin.  The complex scalar field includes two physical particles, a pseudoscalar $a_\tau$ and a scalar $\varphi_\tau$, which have Yukawa couplings to $\taup$ and $\tau$.

We have shown that the standard decay modes of the Vlepton, $\taup \to  \nu W, \, \tau Z$, or $\tau h^0$, are the main ones when the 
pseudoscalar $a_\tau$ is a pseudo-Nambu-Goldstone boson, while the more exotic  $\taup \to  \tau  \atau$ decay may dominate if vectorlike masses are included in the Lagrangian. Electroweak production of a Vlepton pair can lead to a variety of signals at the LHC, including $\tau^+\tau^- + 4\gamma$, $6\tau$, $4\tau + h^0$, and several others discussed in Section \ref{sec:production}.
Depending on the $a_\tau$ and $\taup$ masses, and on some dimensionless parameters (especially the mixing parameter $s_L$, and the Yukawa coupling $y_\taup$), the decay length of $a_\tau$ in the lab frame (see Figure~\ref{fig:median_decaydist}) may vary from microscopic distances to hundreds of meters.

For a decay length of several meters, which naturally occurs for a range of parameters, the muon system of the detector can be used to search for the energy deposited by the decaying $a_\tau$.
This is true at CMS independent of the decay mode, while at ATLAS it easier to search for $a_\tau \to \tau\tau$ than for 
$a_\tau \to \gamma\gamma$ due to the smaller average density of the detector. As shown in  Figure~\ref{fig:limits_muonsystem}, 
we estimate that a dedicated search at CMS
for a signal with two prompt taus and two pairs of displaced photons (see the diagram in Figure~\ref{fig:diagram4gamma}) would probe Vlepton masses above 600 GeV after Run 3, and above 1 TeV after the high-luminosity runs of the LHC.

Longer decay lengths of $a_\tau$ may be probed at proposed LHC experiments dedicated to LLP searches, like MATHUSLA \cite{Chou:2016lxi} or CODEX-b \cite{Gligorov:2017nwh}.
Shorter $a_\tau$  decay lengths can be probed at ATLAS and CMS with future searches for displaced  collimated photons in the electromagnetic calorimeter, using again the prompt taus as triggers. 
In fact, even for the range of  $a_\tau$ decay lengths most suitable for an analysis of the energy deposited in the muon system, like the one discussed in Section \ref{sec:displaced}, the sensitivity would be improved by including events where at least one $a_\tau$ decays in the calorimeter.


\bigskip\bigskip\bigskip\bigskip\bigskip

\noindent
{\it Acknowledgments:}  We thank Artur Apresyan, Patrick Fox, Ka Hei Martin Kwok, Cristi\'an Pe\~na, Si Xie, and Felix Yu for discussions.
Fermilab is administered by Fermi Research Alliance, LLC under Contract No. DE-AC02-07CH11359 with the U.S. Department of Energy, Office of Science, Office of High Energy Physics.
  
  \newpage
    
\providecommand{\href}[2]{#2}\begingroup\raggedright

\vfil

\begin{thebibliography}{10}  
  
\addcontentsline{toc}{section}{References}  
    
    
\bibitem{ATLAS:2018ziw}
 CMS Collaboration,
``Search for pair production of vector-like quarks in leptonic final states in proton-proton collisions at $\sqrt{s} = \! 13$ TeV,'' [arXiv:2209.07327 [hep-ex]].  \\ 
 M.~Aaboud \textit{et al.} [ATLAS],
``Combination of the searches for pair-produced vector-like partners of the third-generation quarks at $\sqrt{s} =$ 13 TeV,"   
Phys. Rev. Lett. \textbf{121} (2018) no.21, 211801
[arXiv:1808.02343 [hep-ex]].
        
    
\bibitem{CMS:2022nty}
A.~Tumasyan \textit{et al.} [CMS],
``Inclusive nonresonant multilepton probes of new phenomena at $\sqrt{s}$ = 13 TeV,''
[arXiv:2202.08676 [hep-ex]].
    
\bibitem{CMS:2019hsm}
A.~M.~Sirunyan \textit{et al.} [CMS],
``Search for vector-like leptons in multilepton final states," 
Phys. Rev. D \textbf{100} (2019) no.5, 052003
[arXiv:\href{https://arxiv.org/pdf/1905.10853.pdf}{1905.10853} [hep-ex]].

\bibitem{Muse:2022xgu}
J.~Muse,
``Search for third generation vector-like leptons with the ATLAS detector,'' University of Oklahoma PhD Thesis, March 2022,
\href{https://inspirehep.net/literature/2052057}{inspirehep.net/literature/2052057}

\bibitem{Kumar:2015tna}
N.~Kumar and S.~P.~Martin,
``Vectorlike leptons at the Large Hadron Collider,''
Phys. Rev. D \textbf{92} (2015) no.11, 115018
[arXiv:1510.03456 [hep-ph]].  \\
P.~N.~Bhattiprolu and S.~P.~Martin,
``Prospects for vectorlike leptons at future proton-proton colliders,''
Phys. Rev. D \textbf{100} (2019) no.1, 015033
[arXiv:1905.00498].

\bibitem{Falkowski:2013jya}
A.~Falkowski, D.~M.~Straub and A.~Vicente,
``Vector-like leptons: Higgs decays and collider phenomenology,''
JHEP \textbf{05} (2014), 092
[arXiv:1312.5329 [hep-ph]].

\bibitem{CMS:2022pvz}
CMS Collaboration,
``Search for pair-produced vector-like leptons in $\geq3\mathrm{b} + \mathrm{N}{\tau}$ final states,''
report PAS-B2G-21-004, March 2022,
\href{http://cds.cern.ch/record/2803736/files/B2G-21-004-pas.pdf}{cds.cern.ch/record/2803736}
    
\bibitem{Dobrescu:2016pda}
B.~A.~Dobrescu and F.~Yu,
``Exotic signals of vectorlike quarks,''
J. Phys. G \textbf{45} (2018) no.8, 08LT01
[arXiv: \href{https://arxiv.org/pdf/1612.01909.pdf}{1612.01909}]. 
        
\bibitem{CMS:2021juv}
A.~Tumasyan \textit{et al.} [CMS],
``Search for long-lived particles decaying in the CMS end cap muon detectors,"  
Phys. Rev. Lett. \textbf{127} (2021) no.26, 261804
[arXiv:\href{https://arxiv.org/pdf/2107.04838.pdf}{2107.04838}].   
  
\bibitem{ATLAS:2018tup}
M.~Aaboud \textit{et al.} [ATLAS],
``Search for long-lived particles produced in $pp$ collisions at $\sqrt{s}=13$  TeV that decay into displaced hadronic jets in the ATLAS muon spectrometer,''
Phys. Rev. D \textbf{99}, no.5, 052005 (2019)
[arXiv:1811.07370 [hep-ex]].
    
\bibitem{ATLAS:2019jcm}
G.~Aad \textit{et al.} [ATLAS],
``Search for long-lived neutral particles produced in $pp$ collisions at $\sqrt{s} = 13$ TeV decaying into displaced hadronic jets in the ATLAS inner detector and muon spectrometer,''
Phys. Rev. D \textbf{101}, no.5, 052013 (2020)
[arXiv:1911.12575 [hep-ex]].
    
\bibitem{Alimena:2019zri}
J.~Alimena \textit{et al},
``Searching for long-lived particles beyond the Standard Model at the Large Hadron Collider,''
J. Phys. G \textbf{47} (2020) no.9, 090501
[arXiv:\href{https://arxiv.org/pdf/1903.04497.pdf}{1903.04497}].

\bibitem{Dobrescu:2021fny}
B.~A.~Dobrescu,
``Quark and lepton compositeness: A renormalizable model,''
Phys. Rev. Lett. \textbf{128} (2022) no.24, 241804
[arXiv:2112.15132 [hep-ph]]. \\     
B.~Assi and B.~A.~Dobrescu,
``Proton decay from quark and lepton compositeness,''
JHEP \textbf{12} (2022), 116
[arXiv:2211.02211 [hep-ph]].    

\bibitem{Dobrescu:2014fca}
B.~A.~Dobrescu and C.~Frugiuele,
``Hidden GeV-scale interactions of quarks,''
Phys. Rev. Lett. \textbf{113} (2014), 061801
[arXiv:1404.3947 [hep-ph]].


\bibitem{Bell:2019mbn}
N.~F.~Bell, M.~J.~Dolan, L.~S.~Friedrich, M.~J.~Ramsey-Musolf and R.~R.~Volkas,
``Electroweak baryogenesis with vector-like leptons and scalar singlets,''
JHEP \textbf{09} (2019), 012
[arXiv:1903.11255 [hep-ph]].
 
\bibitem{Endo:2014hza}
M.~Endo and T.~Yoshinaga,
``Lepton universality test of extra leptons using hadron decay,''
[arXiv:1404.4498 [hep-ph]].

\bibitem{Dobrescu:2009vz}
B.~A.~Dobrescu, K.~Kong and R.~Mahbubani,
``Prospects for top-prime quark discovery at the Tevatron,''
JHEP \textbf{06} (2009), 001
[arXiv:0902.0792 [hep-ph]].
        
\bibitem{ATLAS:2020xea}
G.~Aad \textit{et al.} [ATLAS],
``Test of the universality of $\tau$ and $\mu$ lepton couplings in $W$-boson decays,''
Nature Phys. \textbf{17} (2021) no.7, 813-818
[arXiv:2007.14040]. 
    
\bibitem{ParticleDataGroup:2022pth}
R.~L.~Workman \textit{et al.} [Particle Data Group],
``Review of Particle Physics,''
PTEP \textbf{2022} (2022), 083C01

\bibitem{Han:2003wu}
T.~Han, H.~E.~Logan, B.~McElrath and L.~T.~Wang,
``Phenomenology of the little Higgs model,''
Phys. Rev. D \textbf{67} (2003), 095004
[arXiv:hep-ph/0301040 [hep-ph]].  \\
M.~Perelstein, M.~E.~Peskin and A.~Pierce,
``Top quarks and electroweak symmetry breaking in little Higgs models,''
Phys. Rev. D \textbf{69} (2004), 075002
[arXiv:hep-ph/0310039 [hep-ph]].

\bibitem{Chanowitz:1985hj}
M.~S.~Chanowitz and M.~K.~Gaillard,
``The TeV physics of strongly interacting W's and Z's,''
Nucl. Phys. B \textbf{261} (1985), 379-431


\bibitem{Alwall:2014hca}
J.~Alwall \textit{et al.}, 
``The automated computation of tree-level and next-to-leading order differential cross sections, and their matching to parton shower simulations,''
JHEP \textbf{07}, 079 (2014)
[arXiv:1405.0301 [hep-ph]].

\bibitem{Alloul:2013bka}
A.~Alloul, N.~D.~Christensen, C.~Degrande, C.~Duhr and B.~Fuks,
``FeynRules  2.0 - A complete toolbox for tree-level phenomenology,''
Comput. Phys. Commun. \textbf{185}, 2250-2300 (2014)
[arXiv:1310.1921 [hep-ph]].

\bibitem{Ball:2013hta}
R.~D.~Ball \textit{et al.} [NNPDF],
Nucl. Phys. B \textbf{877}, 290-320 (2013)
doi:10.1016/j.nuclphysb.2013.10.010
[arXiv:1308.0598 [hep-ph]].

\bibitem{Sjostrand:2014zea}
T.~Sj\"ostrand \textit{et al.},  ``An introduction to PYTHIA 8.2,''
Comput. Phys. Commun. \textbf{191}, 159-177 (2015)
[arXiv:1410.3012 [hep-ph]].

\bibitem{deFavereau:2013fsa}
J.~de Favereau \textit{et al.} [DELPHES 3],
``DELPHES 3, A modular framework for fast simulation of a generic collider experiment,''
JHEP \textbf{02}, 057 (2014)
[arXiv:1307.6346]. 

\bibitem{Ajjath:2023ugn}
A.~H.~Ajjath, B.~Fuks, H.~S.~Shao and Y.~Simon,
Phys. Rev. D \textbf{107}, no.7, 075011 (2023)
doi:10.1103/PhysRevD.107.075011
[arXiv:2301.03640 [hep-ph]].


\bibitem{CMS:2008xjf}
S.~Chatrchyan \textit{et al.} [CMS],
``The CMS Experiment at the CERN LHC,''
JINST \textbf{3}, S08004 (2008)

\bibitem{ATLAS:2008xda}
G.~Aad \textit{et al.} [ATLAS],
``The ATLAS Experiment at the CERN Large Hadron Collider,''
JINST \textbf{3}, S08003 (2008)

\bibitem{Cottin:2022nwp}
G.~Cottin \textit{et al.}, 
``Long-lived heavy neutral leptons with a displaced shower signature at CMS,''
JHEP \textbf{02} (2023), 011
[arXiv:2210.17446]. 
A.~Mitridate \textit{et al.}, 
``Energetic long-lived particles in the CMS muon chambers,''
[arXiv:2304.06109].

 \bibitem{hepdata.104408}
{CMS Collaboration}, ``{Search for long-lived particles decaying in the CMS
	endcap muon detectors in proton-proton collisions at $\sqrt{s} = $ 13 TeV}.''
{HEPData (collection)}, 2021.
\newblock \url{https://doi.org/10.17182/hepdata.104408}.

\bibitem{triggers}
S. Xie, private communication. \\
G.~Bagliesi [CMS],
``Reconstruction and identification of tau decays at CMS,''
J. Phys. Conf. Ser. \textbf{119}, 032005 (2008). \\
V.~Khachatryan \textit{et al.} [CMS],
``The CMS trigger system,''
JINST \textbf{12}, no.01, P01020 (2017)
[arXiv:1609.02366 [physics.ins-det]].

\bibitem{CMS:2021woq}
 CMS Collaboration,
``Search for direct pair production of supersymmetric partners to the $\tau$ lepton in the all-hadronic final state,"  
report PAS-SUS-21-001,  July 2021,
\href{http://cds.cern.ch/record/2777046/files/SUS-21-001-pas.pdf}{cds.cern.ch/record/2777046}


\bibitem{CMS:2019zxa}
A.~M.~Sirunyan \textit{et al.} [CMS],
``Search for long-lived particles using delayed photons in proton-proton collisions at $\sqrt{s}=$ 13 TeV,''
Phys. Rev. D \textbf{100}, no.11, 112003 (2019)
[arXiv:1909.06166 [hep-ex]].


\bibitem{ATLAS:2015rsn}
G.~Aad \textit{et al.} [ATLAS],
``Search for new phenomena in events with at least three photons collected in $pp$ collisions at $\sqrt{s}$ = 8 TeV,''
Eur. Phys. J. C \textbf{76}, no.4, 210 (2016)
[arXiv:1509.05051 [hep-ex]].

\bibitem{ATLAS:2018dfo}
M.~Aaboud \textit{et al.} [ATLAS],
``A search for pairs of highly collimated photon-jets in $pp$ collisions at $\sqrt{s}$ = 13 TeV,'' 
Phys. Rev. D \textbf{99}, no.1, 012008 (2019)
[arXiv:1808.10515 [hep-ex]].

\bibitem{Chou:2016lxi}
J.~P.~Chou, D.~Curtin and H.~J.~Lubatti,
``New detectors to explore the lifetime frontier,''
Phys. Lett. B \textbf{767} (2017), 29-36
[arXiv:1606.06298 [hep-ph]].

\bibitem{Gligorov:2017nwh}
V.~V.~Gligorov, S.~Knapen, M.~Papucci and D.~J.~Robinson,
``Searching for Long-lived Particles: A Compact Detector for Exotics at LHCb,''
Phys. Rev. D \textbf{97} (2018) no.1, 015023
[arXiv:1708.09395 [hep-ph]].
    
\end{thebibliography}
\end{document}